\documentclass{article}

\usepackage{arxiv}

\usepackage[utf8]{inputenc} 
\usepackage[T1]{fontenc}    
\usepackage{hyperref}       
\usepackage{booktabs}       
\usepackage{amsfonts}       
\usepackage{nicefrac}       
\usepackage{microtype}      
\usepackage{lipsum}
\usepackage{graphicx}
\usepackage{cite}
\usepackage{makecell}
\usepackage{url}            
\usepackage{bm}
\usepackage{amsmath}
\usepackage{subfig}
\usepackage[usenames,dvipsnames]{xcolor}
\usepackage{csquotes}

\usepackage{subfiles}

\newcommand{\new}[1]{\textcolor{black}{#1}}

\newcommand{\beginsupplement}{%
        \setcounter{table}{0}
        \renewcommand{\thetable}{S\arabic{table}}%
        \setcounter{figure}{0}
        \renewcommand{\thefigure}{S\arabic{figure}}%
     }

\title{Impact and dynamics of hate and counter speech online}

\author{%
  Joshua Garland\thanks{Denotes equal contribution. }\\
  Santa Fe Institute\\
  1399 Hyde Park Road, \\
  Santa Fe, NM 87501 USA \\
  \texttt{joshua@santafe.edu} 
  \And
Keyan~Ghazi-Zahedi$^*$ \\
 Max Planck Institute for Mathematics in the Sciences, \\ Inselstrasse 22, 04103 \\Leipzig, Germany \\
 \AND
 Jean-Gabriel Young\\
 Department of Computer Science,\\
 Vermont Complex Systems Center,\\
 University of Vermont, \\ Burlington, VT 05405, USA
\And
Laurent~H{\'e}bert-Dufresne\\
Department of Computer Science,\\
Vermont Complex Systems Center,\\
University of Vermont, \\ Burlington, VT 05405, USA
\And
   Mirta~Galesic \\
   Santa Fe Institute \\
   1399 Hyde Park Road,\\
   Santa Fe, NM 87501 USA \\
   \texttt{galesic@santafe.edu}
}

\begin{document}
\maketitle

\begin{abstract} 
Citizen-generated counter speech is a promising way to fight hate speech and promote peaceful, non-polarized discourse. However, there is a lack of large-scale longitudinal studies of its effectiveness for reducing hate speech. To this end, we perform an \new{exploratory analysis} of the effectiveness of counter speech using several different macro- and micro-level measures to analyze 180,000 political conversations that took place on German Twitter over four years. We report on the dynamic interactions of hate and counter speech over time and provide insights into whether, as in `classic' bullying situations, organized efforts are more effective than independent individuals in steering online discourse. Taken together, our results build a multifaceted picture of the dynamics of hate and counter speech online. \new{While we make no causal claims due to the complexity of discourse dynamics, our findings suggest that organized hate speech is associated with changes in public discourse and that counter speech---especially when organized---may help curb hateful rhetoric in online discourse.}

\end{abstract}

\section{Introduction}
\label{sec:intro}

Hate speech is rampant on many online platforms and manifests in many different forms, e.g., insulting or intimidating, encouraging exclusion, segregation, and calls for violence, as well as spreading harmful stereotypes and disinformation about a group of individuals based on their race, ethnicity, gender, creed, religion, or political beliefs~\cite{bakalis2015,blaya2019cyberhate,gagliardone2015countering,hawdon2017exposure,muller2019fanning,oksanen2018perceived,weber2009manual,youtube_guide,twitter_guide,facebook_guide}. While it is widely accepted that hate speech is a growing problem on online platforms, what to do about it is a point of contention. One proposal is to more or less automatically detect and remove hateful content. This approach would have to overcome several challenges, however, including the nuanced and constantly evolving nature of hate speech, societal and legal norms about free speech, and the possibility of merely moving hate to other platforms rather than eliminating it~\cite{chandrasekharan2017you}.

An emerging alternative is citizen-driven \emph{counter speech}, whereby users of online platforms themselves generate responses to hateful content in order to stop it, reduce its consequences, discourage it, as well as to support the victim and fellow counter speakers ~\cite{benesch2016considerations,rieger2018hate,wachs2019associations,cathyCounterReview2019}, and ultimately increase civility and deliberation quality of online discussions~\cite{ziegele2018journalistic,habermas2015between}. Like hate, counter speech can take many forms, including providing facts, pointing to logical inconsistencies in hateful messages, attacking the perpetrators, supporting the victims, spreading neutral messages, or flooding a discussion with unrelated content~\cite{brassard2018impact,burnap2015detecting,burnap2016us,macavaney2019hate,ribeiro2018characterizing,zhang2019hate,zimmerman2018improving,keller2020combatting,cathyCounter2020}.

The effectiveness of counter-speech in curbing the spread of hatred online is not well understood. In fact, until recently, there had been a lack of longitudinal, large-scale studies of counter speech~\cite{gaffney2019cyberbullying,gagliardone2015countering}. One reason had been the difficulty of designing automated algorithms for discovering counter speech in large online corpora. Past studies have provided insightful analyses of the effectiveness of counter speech, but were limited in scope as they relied on relatively small, hand-coded examples of discourse~\cite{mathew2018analyzing,mathew2019thou,stroud2015changing,ziegele2018journalistic,wright2017vectors, keller2020combatting}. Recently, however,  two self-labeling groups engaged in organized online hate and counter speech on German twitter. One, called Reconquista Germanica (RG), aimed to spread hate and disinformation about immigrants, while the other, called Reconquista Internet (RI), tried to actively resist this discourse using organized counter speech (see Section~\ref{subsec:RGRI} for more details and \cite{keller2020combatting} for a qualitative analysis and in-depth description of the two groups). 

In previous work, we used the existence of these two groups to develop an automated classifier that could identify typical speech from these groups on a very large textual corpus~\cite{garland2020countering}---orders of magnitude bigger than any other study of online hate and counter speech to date. This classification system achieved high accuracy scores on out-of-sample data and was also verified against human judgment. Here, we use this classification system to perform a longitudinal study on more than 180,000 fully-resolved (out-of-sample) Twitter conversations that took place from 2015 to 2018 within German political discourse to study the \emph{dynamics} of hate and counter speech and \new{to gain descriptive insight} into the potential effectiveness of counter speech.

We anticipate that both organized counter and hate speech will be more effective than independent individual efforts. We develop this theoretical expectation based on two relevant lines of literature. One is the literature on bullying, a phenomenon that shares some of its causes and manifestations with hate speech \cite{blaya2019cyberhate}. The presence of peers is important for both bullying and bully-opposing behaviors. Bullies often seek an approving crowd, because crowd's attention justifies their behavior and helps stifle resistance, while also enabling them to achieve visibility and social status they seek~\cite{salmivalli2014participant}. Bystanders, as well, often look to others when deciding whether to actively oppose the bullying and help the victim. The presence of such peers increases the chance that an opposing individual will receive peer support and protection~\cite{gini2008determinants,salmivalli2011bystanders}. These observations translate to the world of online hate and counter speech, where individuals who engage in either kind of speech can become targets of online hate themselves\cite{cathyCounter2020}, and may even be threatened by physical violence. Seeing that one is a lone countering voice in a flood of similar messages can discourage individuals from engaging in opposing speech, as the effort put in exposing one's own view might seem futile\cite{cathyCounter2020}.

The other line of literature supporting the expectation  that organized efforts will be more effective is the research on social norms. Numerous studies have demonstrated that perceived beliefs and behaviors of others influence people's own political attitudes~\cite{huckfeldt1995citizens,lazer2011networks,sinclair2012social}, university evaluations~\cite{brown2015student}, pro-environmental behaviors~\cite{farrow2017social}, financial decisions~\cite{banerjee2013diffusion}, voting behavior~\cite{bond201261}, food choice~\cite{croker2009social}, and health-related behaviors~\cite{christakis2009connected}. Similarly, the presence of opposing voices in online discussions can be viewed as a signal of how wide-spread these views are in the overall population ~\cite{alvarez2018normative, matias2019preventing}. These perceptions can guide people's reactions to and acceptance of hate and counter discourse. When opposition is organized so that proponents of a particular view come together to post comments in the same discussion (a behavior that is one of the basic operating principles for both RG and RI), this view will become more visible to others. The resulting change in descriptive norm can encourage others with similar views to become more vocal about their positions, further reinforcing the norm. On the other hand, this heightened visibility of a specific discourse (hate or counter) can also mobilize the opposing group to post more countering comments~\cite{cathyCounter2020}. 

\new{While we cannot make causal claims about the impact these groups may have had on the broader German society or vice versa, our data provides a unique opportunity to perform an exploratory analysis of the interplay of organized hate and counter speech in a longitudinal online setting. }

\begin{figure}%
   \centering
   \includegraphics[width=0.3\textwidth]{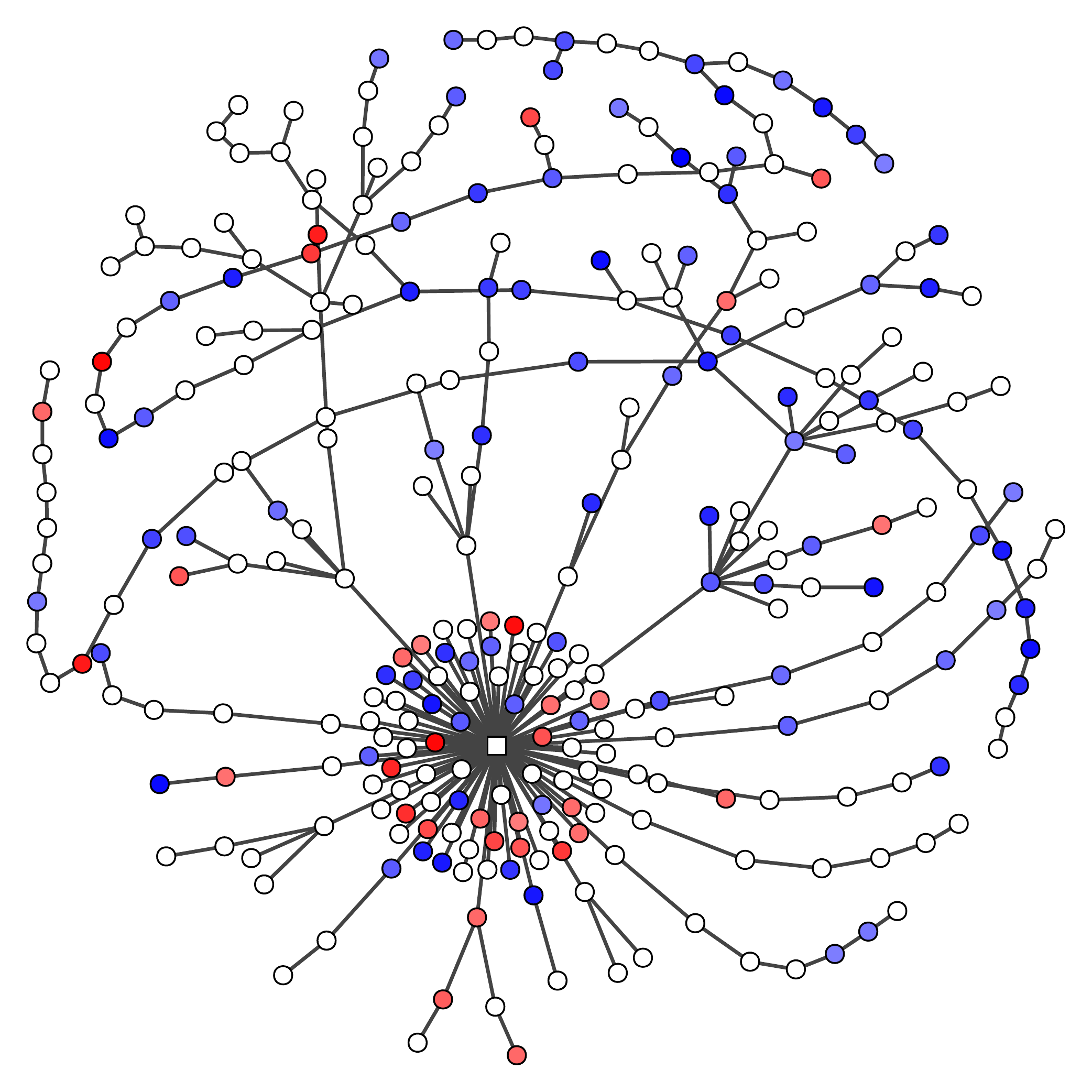} 
    \includegraphics[width=0.3\textwidth]{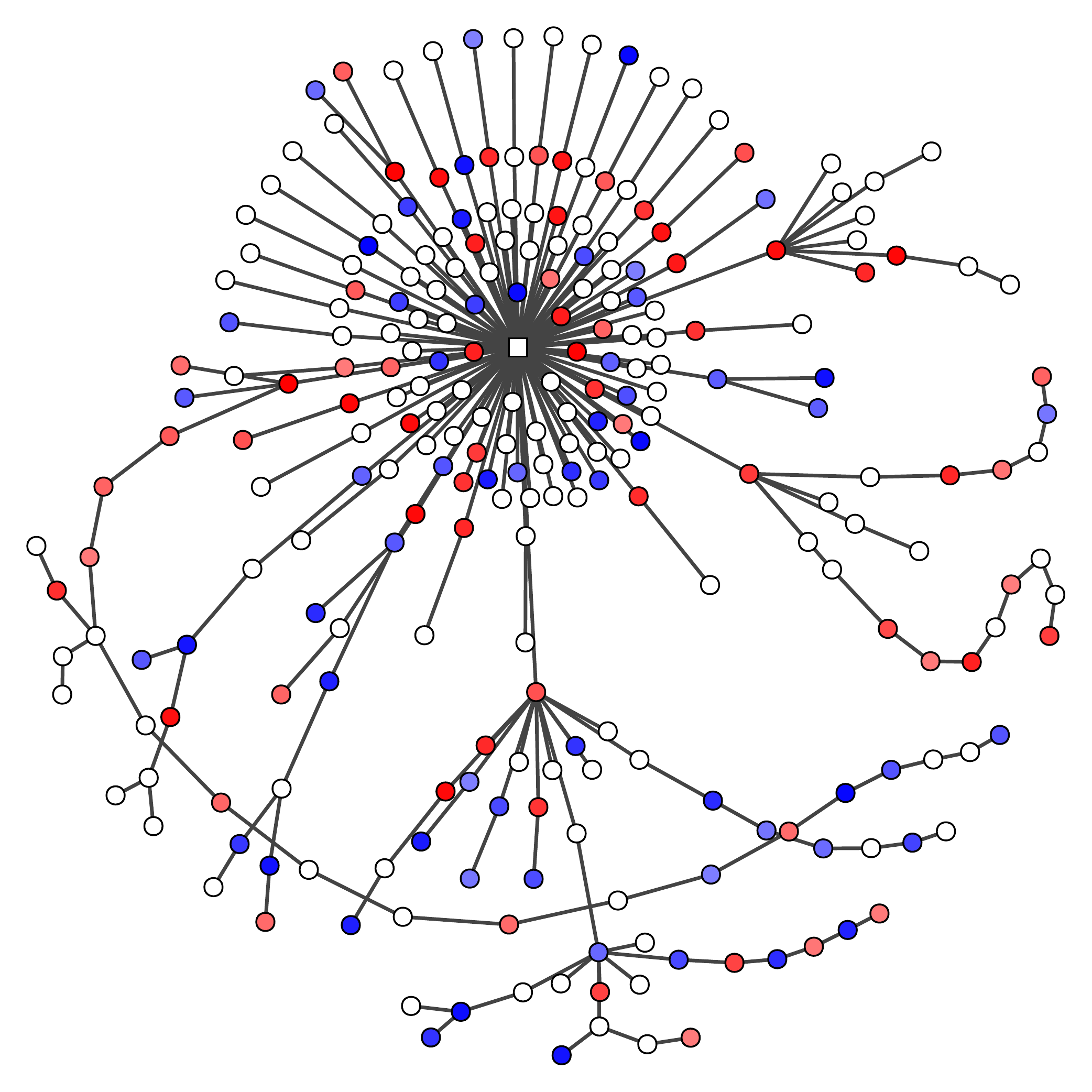}
    \includegraphics[width=0.3\textwidth]{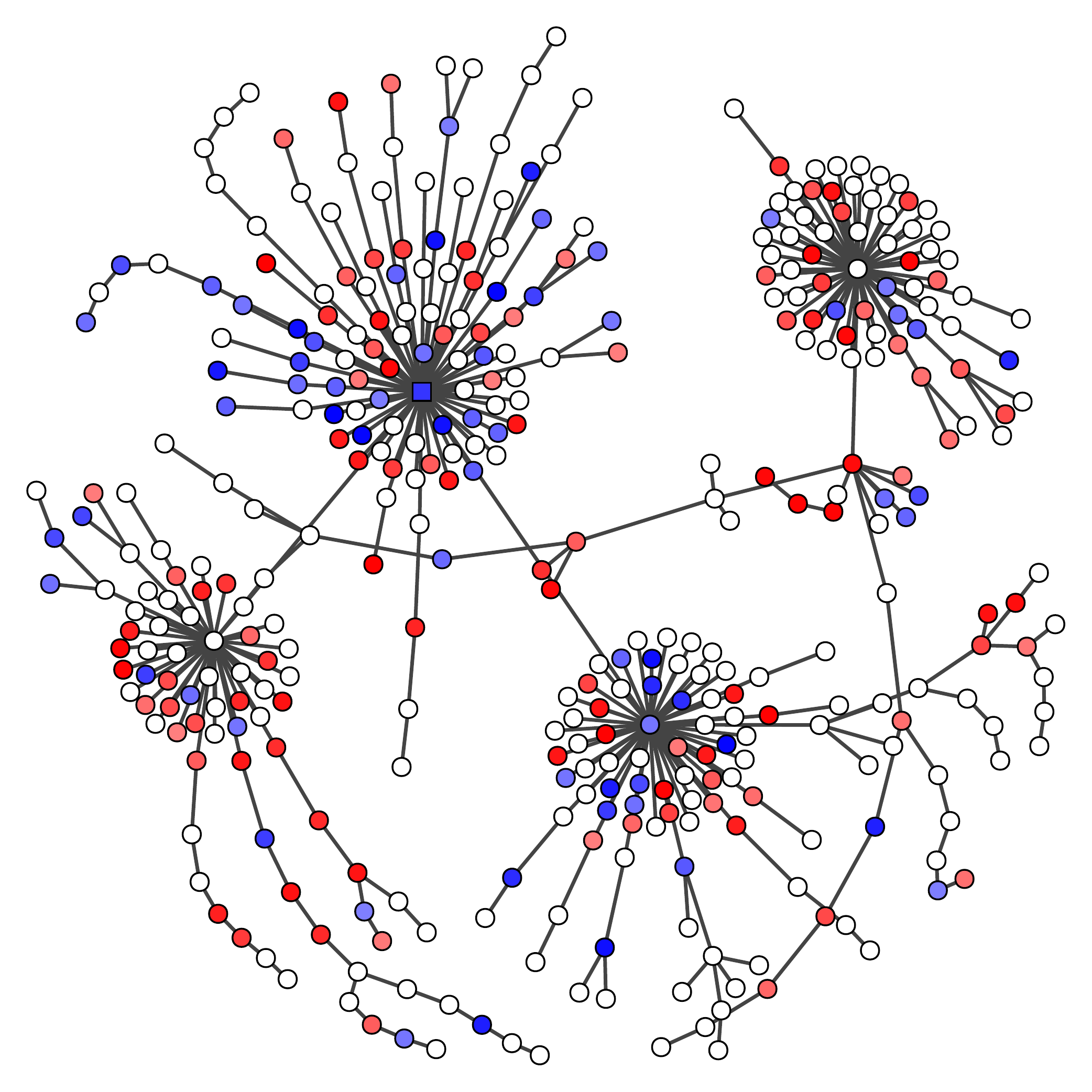}
   \caption{\textbf{Examples of reply trees (Twitter conversations).} Nodes are tweets and edges denote a ``replied to" relationship. Node colors denote whether the content of a tweet was hate (red), counter (blue) or neutral speech (white), with intensity of colors representing the magnitude of the hate score according to the classifiers presented in~\cite{garland2020countering}. %
   }%
   \label{fig:exampletrees}%
\end{figure}

\section{Background, Data and Methods}
\label{sec:methods}

\subsection{Background on Reconquista Germanica (RG) and Reconquista Internet (RI)}
\label{subsec:RGRI}
Here we provide a brief overview of the two focal groups involved in our case study. For a curious reader we recommend Ref.~\cite{keller2020combatting} for an in-depth analysis of both groups.  Reconquista Germanica (RG) is a highly organized far-right troll army which officially formed two months prior to the German federal election with the purpose to influence the election to benefit Germany's radical-right party the Alternative f{\"u}r Deutschland  (AfD) in 2017. However, recruitment for RG and related hate groups started earlier in late 2016 \cite{davey2017fringe}. RG was organized hierarchically, and prospects had to prove their political views in a series of online interviews, before they were granted access to the Discord server. Once admitted, people could climb the ranks by proving themselves in coordinated actions, which were given in daily orders to the entire group. Actions were then coordinated between multiple social media outlets and each member was usually responsible for multiple accounts. It is unclear how many active members RG had and how many accounts they controlled, but at its peak, RG might have had around 5,000 active members \cite{keller2020combatting}.

Reconquista Internet (RI) was established in late April 2018 shortly after the exposure of RG by a group of journalists~\cite{tag-rg,funk-docu}, upon the encouragement of German TV host Jan Böhmermann during the popular German satire news show Neo Magazin Royal. While RI was originally a satirical response to RG it quickly grew to be the largest counter speech group to date in Germany~\cite{keller2020combatting}. RI was fully functional by May 2018. RI’s goal were to spread positive messages, restore civil discourse and to directly counter the hate being spread by RG with ``love and reason~\cite{RICodex}." Like RG, RI was organized into groups which formed around specific social media platform, i.e., there was a group for Twitch, Facebook, Twitter, and so on~\cite{RICodex}. Each group had two leaders, who organized activities within their group. To communicate with one another, call for help etc., several Discord channels were maintained, e.g., a channel for sharing links to tweets which were selected for engagement, under attack etc. At its peak around May 10, 2018, RI had 62,000 registered users on its Discord server. This number quickly decreased to 4-5,000 active members during the first few months, followed by further splintering into smaller independent groups by the end of July 2018. \new{The details presented here about the internal workings, organization and strategies of RI are a synthesis of private correspondences with active members of RI, as well as Reconquista Internet's Codex \cite{RICodex} and finally the information presented about RI in Ref.~\cite{keller2020combatting}.}

One difference between these two groups was their organizational structure and tactics. As mentioned previously, RG, for example, would take part in highly-coordinated cyberhate attacks where they were instructed who to attack and how. With instructions\new{~\cite{tag-rg,funk-docu}} like (translated from German): \emph{Young women who came straight from university make classic victims...Offend. And then draw every register. Do not leave anything out. Weak point are often the family. Always have a repertoire of insults that you can adapt to the respective opponent...As soon as you see them (media/politicians) spilling their lies and their poison into the world again, tell them our opinion, engage them in discussions, mark their lies as \#fakenews and troll the **** out of them.}  In contrast, RI was more loosely organized\cite{keller2020combatting}. Instead of giving direct orders of who/what to respond to and how, RI responded to calls for protection via Discord channels, and instead of being told what to say, RI followed a few general principles when responding, outlined in their codex\cite{RICodex}, e.g, \emph{Human dignity is inviolable. We are not against `them'. We want to solve problems together and act guided by mutual respect, love, and reason, We do not wage war but seek conversation.} 

While both RG and RI were active on several social media platforms, for our analysis, we focused on a sample of their Twitter presence, \new{focused around several prominent German news organizations and public figures} which enabled us to study the structure of the resulting conversations and the dynamic interplay between the two groups over time.

\subsection{\new{Data}}
\new{For the analysis reported in this manuscript we performed two independent data collection phases:}
\begin{enumerate}
    \item \new{\textbf{Classifier Training Data Collection:} We collected millions of Tweets originating from approximately 3,700 known RG and RI members to train a classification system to identify hate and counter speech typical of these groups, as well as neutral speech not typical of either group. This data is described in Section~\ref{sec:TTdata} and the classification system as well as its accuracy and verification is described in \cite{garland2020countering} and Section~\ref{sec:classification}.}
    \item \new{\textbf{Reply Tree Data Collection:} We collected a longitudinal sample of German political conversations (or ``Reply Trees") over a 4-year period during which RG and/or RI were active. For this we specifically targeted root accounts (initiators of conversations) where organized activity regularly occurred by both RG and RI. We collected trees before, during and after these groups were active, to study if there were quantifiable differences in overall political discourse. Details about the collection of this data and choices that were made are discussed in Section~\ref{sec:RTdata}. } 
\end{enumerate} 
\new{Table~\ref{tab:data} summarizes the two resulting data sets. The following two sections provide details on how we went about collecting this data.}
\begin{table}
   \centering
   \begin{tabular*}{\textwidth}{c@{\extracolsep{\fill}}cc}
      \toprule
      \multicolumn{3}{c}{Summary of Classifier Training Data} \\
      \midrule
        Category   & Members Identified & Total Tweets Collected   \\
     \midrule
     Reconquista Germanica (Hate)  & 2,120 & 4,689,294   \\
     Reconquista Internet (Counter) & 1,575 & 4,323,881  \\
  \bottomrule
        \end{tabular*}\newline\newline
   \begin{tabular*}{\textwidth}{c@{\extracolsep{\fill}}cccc}
      \toprule
         \multicolumn{5}{c}{Summary of Reply Tree Data} \\
         \midrule
           Category & Total Trees  & Total Tweets in Trees  & Total Root Accounts & Time Frame   \\
        \midrule
        Complete & 203,711 & 1,649,137 & 9,933 & Jan. 2013 - Dec. 2018 \\
        Filtered & 181,370 & 1,222,240 & 23 & Jan. 2015 - Dec. 2018 \\
    \bottomrule
   \end{tabular*} 
   \vspace{0.5\baselineskip}
   \caption{\new{A summary of all data that was used in this manuscript. The top table summarizes the tweets that were used in \cite{garland2020countering} to train the classifier used in this manuscript. This is broken into two categories: Reconquista Germanica and Reconquista Internet and summarizes how many members were identified from each group and how many tweets were collected from those respective groups. The top table summarizes the Reply Tree Data, including the total number of trees, the total tweets among all trees, the total number of accounts which served as root nodes and the time frame when these reply trees occurred. } }
   \label{tab:data}
\end{table}

\subsubsection{Classifier Training Data Collection}
\label{sec:TTdata}
\new{To train our classification algorithm, we collected a balanced corpus of more than 9 million relevant tweets originating from known RG accounts (4,689,294 tweets) or RI accounts (4,323,881 tweets). For a full discussion of how we identified these members see~\cite{garland2020countering}. But briefly, for RG members we used the list of hate accounts known as the ``B\"ohmermann Liste," which was promoted as a list of accounts that spread hateful rhetoric, promote radical-right propaganda, or engage in hateful speech. This was a list compiled by investigative journalists \cite{funk-docu}, promoted by Jan B\"ohmermann and verified by several RI members. For RI we began with a hand-curated list of 103 accounts comprised of the core RI Twitter team which we received from RI directly. We then expanded this set by looking at the ego network of these 103 core RI members using the Twitter API and then only included users that appeared in at least 5 of the ego networks of core RI members. This list was then further reduced by analyzing their screen names and bios for known RI badges (see~\cite{garland2020countering} for more details). Tweets that came from RG accounts were labeled as ``hate speech" and tweets that came from RI accounts were labeled as ``counter speech." This data allowed us to build a classifier to identify speech similar to that of RG and RI members. We discuss the ramifications of this in more detail in Section~\ref{sec:classification} and we also point the reader to~\cite{garland2020countering} for a complete description on exactly how these tweets were selected, processed and used. }

\subsubsection{Reply Tree Collection}
\label{sec:RTdata}

\new{To perform an exploratory analysis of the potential impacts of organized hate and counter speech on German political discourse over time it was necessary to collect a longitudinal dataset of German political conversations. This collection involved three main challenges: selecting appropriate conversation initiators (``root accounts") to monitor over time, selecting an appropriate time frame to monitor the root accounts and finally collecting as many of the reply trees as possible for each root account across the desired time frame.}

\new{The first challenge was to identify a consistent source of conversations that would collectively represent a balanced sample of German political discourse during our time frame of study. Additionally, we wanted to ensure these conversations contained activity by both RG and RI. Hence, we identified conversations by selecting root accounts that had a large number of followers, tweeted frequently, existed for an extended and continuous period of time, and whose tweets and subsequent conversations were political in nature. Perhaps most importantly,  we selected root accounts that we knew, from private correspondences with RI, were often targeted by RG. As a starting point, we focused our data collection efforts on conversations that grew in response to @tagesschau. We chose to start with Tagesschau because it is widely considered as a reliable and balanced news source~\cite{german-news},  has existed for a long  time, and has many followers on Twitter. As our study continued, we expanded the list of root accounts to include more accounts where RG/RI activity was taking place regularly. These included additional mainstream news outlets, journalists and bloggers who were targeted frequently by RG and finally, politicians and other relevant public figures. Each account added was done so based on advice by members of RI as to where the most relevant organized actives were occurring.} 

\new{Our next challenge was to select a time frame to study German political discourse which included the time of RG and RI activity but also included enough of a lead up to these organized groups that we could monitor changes in discourse over time. To this end, we chose to start collecting reply trees from January 2013, the point preceding the founding of the Alternative f\"ur Deutschland (AfD). We monitored the discussions continuously until the end of 2018, when Twitter made fundamental changes to their website which made it impossible to collect trees in the manner we describe below. As such, our dataset ends in December of 2018. Nevertheless, the time frame from January of 2013 to December of 2018 allowed us to study discourse leading up to the formation of both organized groups, including a long stretch of time where both groups existed and interacted.}

\new{Our final challenge was to collect a sample of fully-resolved conversations from these root accounts across our study period. As the original Twitter API did not provide functionality to collect reply trees in their entirety, we developed two types of custom scrapers to collect these reply trees. The first type of scraper identified (at random) a one month contiguous sample of conversations originating from a single root account of interest which was missing from our collection of conversations. For example, these scrapers would randomly select one root account of interest, e.g., @tagesschau, as well as a one month period between January 2013 the present date, e.g., 1st of March 2016 to 31st of March 2016. The scraper would then systematically find all top-level urls\footnote{\new{These urls had the format ~\url{https://www.twitter.com/<screen\_name>/status/<tweet\_id>}, where $<$screen\_name$>$ and $<$tweet\_id$>$ refer to the name of the Twitter account and the unique id of the tweet.}} for tweets by that account in that time period, and stored these in a file. The second type of scraper would then randomly select a top-level url from this file and manually scrape the resulting conversation in its entirety from the top-level urls' raw html. A log of previously scraped reply trees ensured that each tree was only scraped once. The scraper ran continuously from June to December of 2018. In total, we collected 203,711 conversations (reply trees) that grew in response to tweets of 9,933 root accounts between January of 2013 to December of 2018. }

\new{The data gathering process stopped in early 2019 following abrupt changes to Twitter's website at that time.  Our system had no explicit bias or preference as to which reply trees were collected at any given time in the collection process. Consequently, accounts that were added to the list later were sampled less often and, conversely, accounts that were added earlier had a larger number of conversations. As this could potentially create bias in the resulting sample, for the analysis reported here we limited the dataset to 181,370 conversations, containing 1,222,240 tweets, which grew in response to tweets from 23 accounts for which we had continuous coverage throughout the period of January 2015 (the beginning of the migrant crisis in Europe) to December 2018. 
}

\new{The final 23 root accounts included mainstream\footnote{\new{At the time of the study these news organizations appeared in lists of the most important news outlets, e.g.,\cite{german-news}. In addition, we learned from private conversations with RI which of these accounts had organized activity from both sides, which helped us narrow the list to the most relevant news organizations.}} German news organizations (\emph{@diezeit, @derspiegel and @spiegelonline, @faznet, @morgenmagazin, @tagesschau, @tagesthemen, @zdfheute}), well-known journalists and bloggers (\emph{@annewilltalk, @augstein, @dunjahayali, @janboehm, @jkasek, @maischberger, @nicolediekmann}), and politicians ( \emph{viz., @cem\_oezdemir, @c\_lindner, @goeringeckardt, @heikomaas, @olafscholz, @regsprecher, @renatekuenast}). Each of these accounts have witnessed organized activity by either RG or RI or both. 
The number of trees per day from each type of account was stable across the sampled period, see Figure~\ref{fig:trees_day}. In analyses reported below, to ensure the type of root account (e.g., news, journalist, politician) did not bias the results, we investigated the sensitivity of all results to statistical controls for the type of root account, and found that they had no discernible effect. }

\new{While our team had to go to great lengths to obtain the fully-resolved Twitter conversations used for this analysis, Twitter is now making this kind of analysis easier. With the advent of the second version of the Twitter API as well as the academic product track, academic researchers will now be able to  directly obtain fully-resolved historical conversations. However, at the time of submitting this paper, the data that could be extracted from the API is incomplete for most past conversations---and much of the signal of extreme forms of discourse has been deleted due to suspensions, user deleted accounts and user deleted tweets (see Supplementary \ref{sup:academicAPI} for more details).}

\subsection{\new{Classification of Hate, Counter and Neutral Speech}}
\label{sec:classification}

\new{To study how RG and RI may have impacted German  political discourse in our reply tree dataset it was necessary to first build a classification system which could recognize the hateful rhetoric of RG, the counter speech of RI as well as neutral political discourse. In \cite{garland2020countering} we built such a classification system using the data described in Section~\ref{sec:TTdata}. The focus of the present work is applying that system to the dataset described in Section~\ref{sec:RTdata}, but we describe that classification system for completeness. In addition, we discuss the results verifying that this classifier performed as expected e.g., agreeing with human intuition.} 

The classification pipeline we used to classify tweets in reply trees consisted of two stages~\cite{schmidt2017survey}: extraction of features from text and classification using those features. To extract the features, we pre-processed the data and then constructed paragraph embeddings, also known as doc2vec models~\cite{doc2vec}, using the standard gensim implementation~\cite{rehurek_lrec}. We performed a parameter sweep following standard practice and the guidelines of~\cite{lau2016empirical}. This sweep included the analysis of several doc2vec parameters e.g., maximum distance between current and predicted words, ``distributed-memory'' vs ``distributed bag of words'' frameworks, three levels of stop word removal and five different document label types (see~\cite{garland2020countering} for more details). Each version of the doc2vec model was trained on five different but partially overlapping training sets. Each training set included 500,000 randomly selected tweets originating from RG accounts and another 500,000 coming from RI accounts.

These trained doc2vec models allowed us to extract features, i.e., infer a feature vector, from a given tweet. To classify each tweet as either hate or counter we then coupled each doc2vec model with a regularized logistic regression classifier, as implemented by scikit-learn\cite{scikit-learn}. These logistic regression functions were trained on the same training set as the corresponding doc2vec model. 

\new{Note that the doc2vec models and logistic regression functions used only the tweet content. Features such as, badges, screen names etc. were not used in this phase explicitly. However, the initial labeling of the accounts did consider these features, so one could view RG and RI badges as secondary features used by the classification scheme.}
 
By pairing a trained doc2vec model and a logistic regression function, we were able to assign a probability $p_h(t)$ to each tweet that it belongs to either the hate or counter class. We then recoded this probability to a hate score $S_h(t)$ ranging from -1 to 1, using 
\begin{equation}
    S_h(t)=2p_h(t)-1 \label{eq:hate_score}
\end{equation} 
where negative values mean that a tweet is similar to prototypical RI counter speech and positive values mean that it is similar to prototypical RG hate speech. For the results reported here we used an ensemble learning approach where 25 independent doc2vec, logistic regression pairings voted on each tweet in a given tree. The tweet's final hate score $S_h(t)$ was then defined as the average hate score assigned to that tweet across all 25 pairings. 

For final classification of tweets, we used a confidence voting system with thresholding to assign each tweet a label. For the majority of the analyses reported in this paper (unless stated otherwise), if $S_h(t)\ge 0.4$ $t$ was labeled hate, and if $S_h(t)\le -0.4$ $t$ was labeled counter. If $-0.4 < S_h(t)< 0.4$ then the tweet was labeled neutral. These scores effectively mean that the ensemble of classifiers was at least 70\% confident in labeling a tweet as either hate or counter speech.  

Depending on the threshold being used, this classification system achieved F1 scores ranging from $0.762$ to $0.97$ on balanced test sets containing equal proportion of hate and counter speech. See Table~\ref{tab:results} for a full summary of classification performance metrics. For the threshold used in this paper $\gamma=0.40$  the classification pipeline achieved F1 scores of 0.877. This accuracy exceeds previous studies that used smaller unbalanced data sets and achieved F1 scores ranging from 0.49 to 0.77 \cite{mathew2019thou, mathew2018analyzing, ziems2020racism}. 
\begin{table*}
   \centering
   
   \begin{tabular}{ccccc}
      \toprule
         \multicolumn{5}{c}{Accuracy Scores for Classification System} \\
         \midrule
            $\gamma$  & Precision  & Recall & {F1} & Percent Labeled  \\
        \midrule
        0.0 & 0.763 & 0.762 & {0.762} & 100\% \\ 
        0.20 & 0.827 & 0.827 & {0.827} & 76\% \\ 
        \textbf{0.40} & \textbf{0.877} & \textbf{0.876} & \textbf{0.877} & \textbf{57\%} \\  
        0.60 & 0.917 & 0.917 & {0.917} & 41\% \\  
        0.80 & 0.958 & 0.958 & {0.958} & 25\% \\ 
        0.90 & 0.977 & 0.977 & {0.977} & 15\% \\ 
    \bottomrule
      \end{tabular}
   \vspace{0.5\baselineskip}
   
   \caption{\new{Classification scores for the panel of experts used to label the Reply Tree tweets using the top 25 experts. $\gamma$ is a confidence threshold on $S_h(t)$, viz., $|S_h(t)=2p_h(t)-1|\ge \gamma$. ``Percent Labeled'' is the percentage of examples in the test set that were labeled as either hate or counter at a confidence level of $\gamma$. The top row represents traditional accuracy measures, which compare favorably to previous studies that used smaller unbalanced data sets and achieved F1 scores ranging from 0.49 to 0.77\cite{garland2020countering}. The remaining rows are more nuanced as not all examples in the tests sets are being labeled correctly or incorrectly and thus these rows need to be viewed with cautious optimism; see \cite{garland2020countering} for an in-depth discussion of these issues.}}
   \label{tab:results}
\end{table*}

While these F1 scores are impressive compared to other similar studies, a careful reader will notice that our classifier is not actually directly classifying hate and counter speech, but that instead we are using \emph{proxies} for hate and counter speech. Indeed, as we have mentioned in the introduction, classifying hate speech remains a difficult challenge to this day. Instead, because of the way this training data was originally labeled, \new{i.e., as an RG or RI tweet}, we are actually classifying speech that is typical of RG vs. RI members. In a sense, we are trading off between potentially noisy labels and the scale of the labeled training corpus, which comes with some risk. Some of the tweets in the training data set which were labeled as hate or counter might have included some other type of speech. At the same time, using these labels allowed us to conduct analyses on a large scale, capturing a wide breadth of both hate and counter speech patterns. We next describe three steps we took to investigate the validity of our automated classification system. (See Ref.~\cite{garland2020countering} for more details on classification trade offs and the care that was taken in selecting the accounts we used to train the classifiers.)

\begin{figure}%
   \centering
   \includegraphics[width=0.8\textwidth]{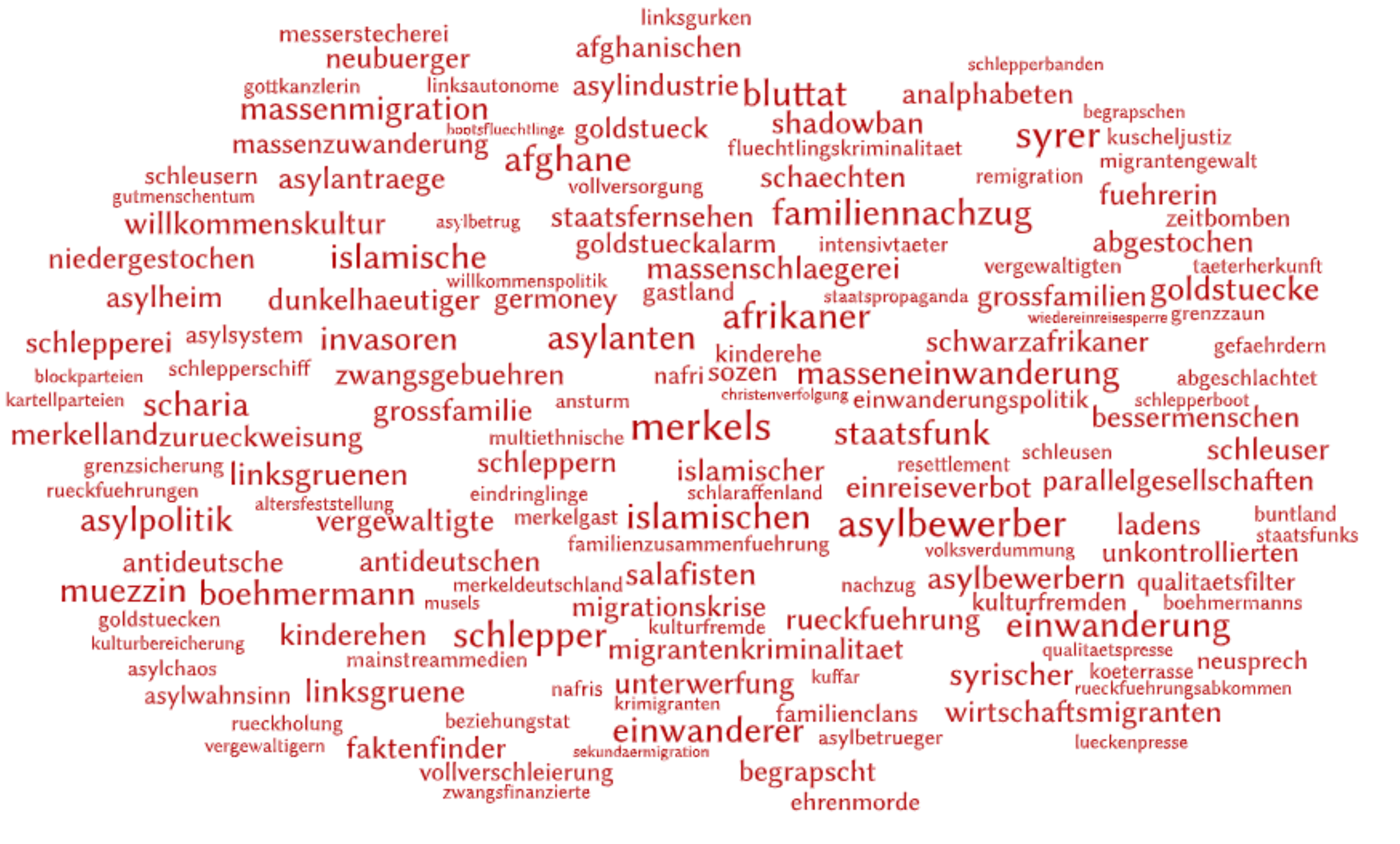}%
   \\
   \includegraphics[width=0.8\textwidth]{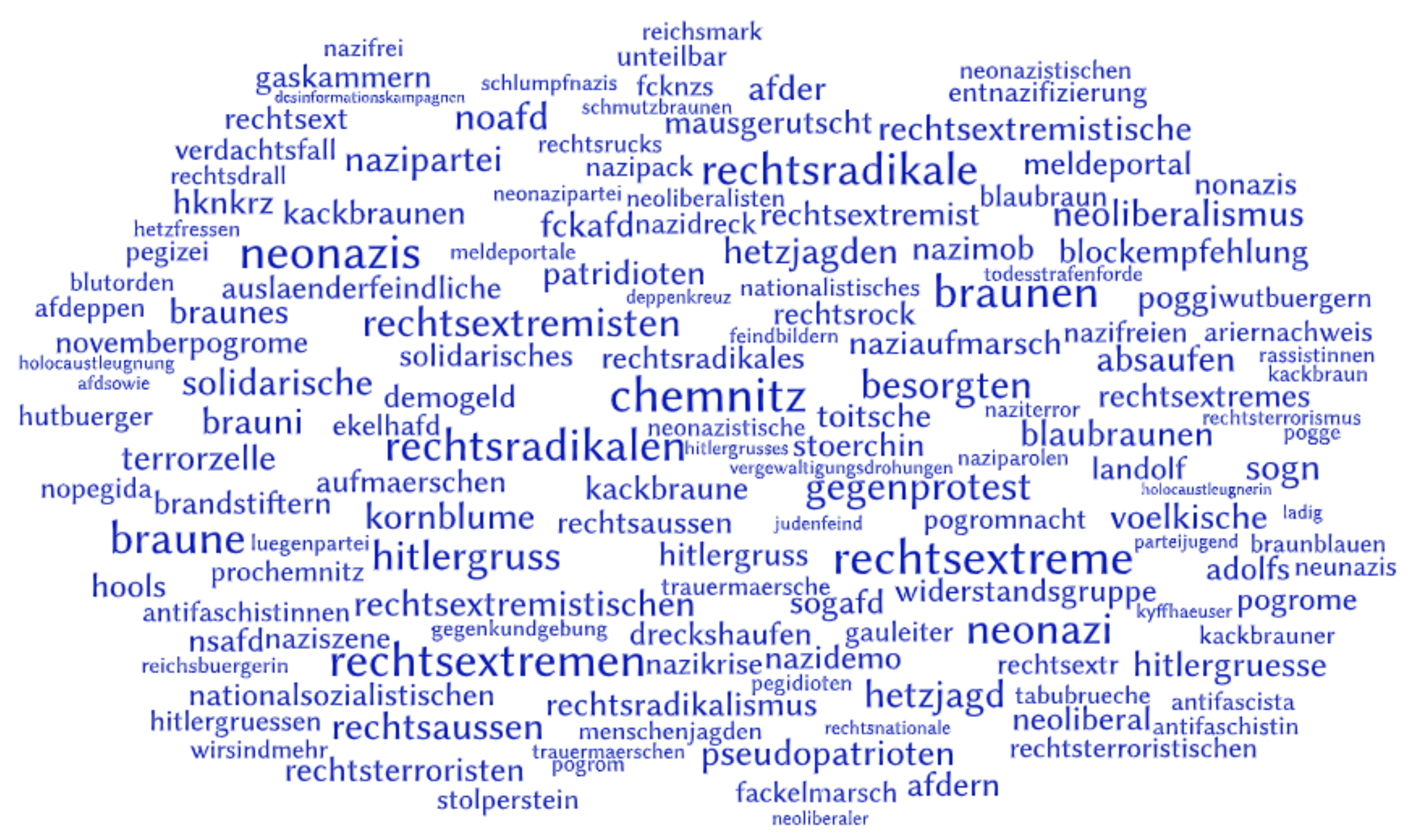}%
   \caption{\textbf{Words that are most frequently used in tweets with high vs. low hate scores.} A: most frequent words appearing in hate tweets that were rarely used in counter tweets. B: The most frequent words used in counter tweets that did not appear in hate tweets.}
   \label{fig:cloud}%
\end{figure}

First, we conducted a qualitative analysis of the speech labeled as hate and counter by our classifier. To this end, we counted the frequency of all tokenized words in the whole corpus, and determined how indicative each word is for tweets labeled as either hate or counter speech~\cite{jaki2019right}. Figure \ref{fig:cloud} shows the top 100 most frequently used words that are specific to each corpus, i.e., these are words that occur the most frequently in hate/counter speech, but which simultaneously occur very infrequently in counter/hate speech. The typical words used by RG (Figure~\ref{fig:cloud} (A)), which we label as hate, focus on Merkel and variations of the word migrant, immigrant, asylum seekers (their typical targets of attack). The typical words used by RI ((Figure~\ref{fig:cloud} (B)) focus on nazis, major neonazi rallies such as Chemnitz and the so-called ``radical right" (Rechtsradikal). All of these tokens align well with our notion of hate and counter speech. 

Second, three of the authors with a working knowledge of the German language read a large sample of classified tweets to confirm that the classifications made sense. We provide a few example tweets typical of RG (hate speech) and RI members (counter speech).  Tweets with high hate scores (typical RG speech) typically discuss purported crimes committed by refugees against Germans, as well as criticize politicians, other public figures, and fellow citizens who are perceived as being too weak or overly welcoming towards refugees and the rising diversity of the German society. Typical examples of these tweets (translated from German) are \emph{``Disgusting: Afghan molested 7 school girls (10-12)  Germany's `cuddling' justice has struck again. A pedophile Afghan that taxpayers have to feed for life was just what was missing in Merkel's `colorful country'!"}
and  \emph{``Call for help from the mother of the raped and murdered 14-year-old Susanna to chancellor Merkel. If Susanna and her mother were refugees, Merkel would have reacted. But Merkel did not say a word about any of the many German girls murdered by refugees.''}

Tweets with high counter scores often point to the similarities of the German radical-right movement and the Nazis, and highlights various ways in which radical-right movements influence German society, calling for counter protests. Typical examples of these tweets (translated from German) are \emph{I find it very deceiving when voters of the \#AFD claim they are not Nazis. Whoever goes to vote should find out more about that party. AFD voters vote for Nazis and thus a right-wing extremist Nazi party. Distancing is not possible. Nazis thrive on mindless followers/supporters.}
and \emph{AfD marches with neo-Nazis and right-wing radicals through Chemnitz, Maaßen, Seehofer and Kretschmer make all migrants a problem and relativize mobs and hunts [targeting refugees]. In this way the hatred creeps from the right towards the middle and becomes normal.}
Further qualitative analysis of a subset of discussions between RG and RI can be found in \cite{keller2020combatting}. 

Third, to verify that our potentially noisy classifier did indeed accurately classify hate and counter speech we conducted a crowdsourcing study to check whether the hate scores derived through this automated pipeline corresponded to human judgment. In our study, human judges evaluated some of the same tweets evaluated by the automated system \new{using the content of the tweets and nothing else}. We selected 28 German-speaking raters to evaluate 5000 randomly selected tweets evenly spread across the whole range of scores $-1\le S_h(t)\le 1$. Raters ranked tweets on a scale of 1 to 5, from ``very likely counter speech'' to ``very likely hate speech,'' with 3 corresponding to neutral content. Each tweet was evaluated by at least 2 different raters(see \cite{garland2020countering} for details). The results suggest that classifier hate scores aligns well with human judgment, with overall correlation between hate scores and human judgments being $r=0.94$. The correlation was somewhat lower for tweets classified as counter speech ($r=0.75$) than for those classified as hate ($r=0.96$). As expected, classifier scores around 0.5 received intermediate hate scores from human judges. \new{It should be noted that the labels assigned by these human classifiers were not used in the labeling process at all. Rather these labelings were only used to validate the classification system's output. The only thing used to label the tweets was the group affiliation of the tweet, i.e., tweeted by  an RG member or coming from an RI member and nothing else.}

\subsection{Statistical Analyses}
\label{subsec:statan}
To investigate whether counter speech had an effect on various indicators of discourse dynamics \new{(such as relative prevalence of hate and counter speech over time, impact of hate and counter speech on subsequent discourse, as well as likes of and discussions initiated by tweets containing hate and counter speech)}  one would ideally like to perform some flavor of statistical causal inference analysis. However, due to a variety of conflating factors, the reply tree time series presented in this paper are difficult to analyze from a causal perspective. The conversations are part of an intricate social system in which it is difficult to define and isolate a reasonable `control group', rendering analyses such as ``difference in differences"~\cite{abadie2005semiparametric} inapplicable in this case. Furthermore, because of unquantifiable outside influences, causal inference methods such as Granger Causality~\cite{granger1969investigating} or Transfer Entropy~\cite{schreiber2000measuring} are of limited value. The lack of evidence for determinism and stationarity in our datasets made state-space~\cite{takens1981detecting} based causal inference, e.g., Convergent Cross Mapping\cite{sugihara2012detecting}, not applicable.

\subsubsection{Change Point Analysis}

While we could not perform rigorous causal inference due to fundamental limitations of the system, we nevertheless put effort into conducting a principled study of the changes in this dynamics over time---even if the analysis could only identify correlations and suggest changes in descriptive statistics such as overall trends. 

In particular, to investigate changes in the proportions of hate and counter speech over time and in association with different events (Figure~\ref{fig:macro_measures}), we conducted both exploratory and confirmatory statistical analyses. Exploratory change point analyses were conducted using the \emph{findchangeptsalgorithm} available in Matlab~\cite{killick2012optimal,aminikhanghahi2017survey}. The algorithm searched for changes in linear trends, always using a standard deviation of the empirical trend as the threshold value. Many change point algorithms have been developed (see \cite{aminikhanghahi2017survey} for a review) to identify changes in either some summary statistic or trends, e.g., mean, variance or information production. However, we were interested in studying changes in overall \emph{trends} in discourse which made \cite{killick2012optimal} an ideal approach. For confirmatory analyses, we tested differences in trends before and after occurrence of Reconquista Germanica and Reconquista Internet.
\subsubsection{Prediction}
To investigate the relationship between the current and future proportion of hate and counter speech, we used a Vector Autoregression (VAR) analysis\cite{var}. For this analysis, we accounted for non-stationarities in each time series in the usual way by taking first differences of the time series shown in Figure~\ref{fig:macro_measures}. Dickey-Fuller tests\cite{dickey1979distribution} showed that the resulting trends were stationary in all three periods. We use these transformed stationary time series for the following VAR analysis.

\subsubsection{Longitudinal mixed linear model}
To investigate the effectiveness of different types of speech to steer a conversation depending on its hate score (Figure \ref{fig:heatmap}), we corrected the impact of focal tweets using a longitudinal mixed linear model. In this model, focal tweet impact was predicted by its score and a host of other factors that could have affected the impact in addition to the score. The focal tweet's position in the tree, tree size, average hate score of the tree, week, and type of Twitter account that started the reply tree (news outlets, politicians, or journalists/bloggers) were included as fixed effects, and variations by tree and week, tree size were allowed by including them as random effects (see Table~\ref{tab:model_diffs}). These corrections did not change the original results much (see top panel of Figure~\ref{fig:heatmap_raw} for uncorrected results). Note that each square in (Figure \ref{fig:heatmap}) represents the average of all focal tweets with a particular hate score posted in a specific week. Different squares include different number of such tweets, as shown in bottom panel of Figure~\ref{fig:heatmap_raw}, and a different number of trees. These and other factors have been statistically accounted for by the model described above. We used similar models to correct all of the other trends shown in the paper. The resulting changes were minimal, so we chose to present the raw data in all other figures.

\subsection{Data Availability Statement}
The models and trends that support the findings of this study are available from the corresponding author upon reasonable request. However, for ethical reasons as well as to protect the identity of counterspeakers raw data will not be released. 

\subsection{Code Availability Statement}
Python and Matlab codes used for different analyses are available from the corresponding author upon request.

\section{Results}
\label{sec:results}

We explore several different macro- and micro-level measures of effectiveness, providing complementary views from different angles. We aim to gain insights into the effectiveness of counter speech by measuring how hate and counter speech interact with each other on macro- and micro-levels. This analysis provides complementary views on effectiveness from different angles. On the \emph{macro-level}, we study changes in the composition of online discourse, reflected in the prevalence of hate, counter, and neutral speech over time. This analysis helps assess the overall deliberation value and civility of discussions~\cite{stroud2015changing,ziegele2016not,ziegele2018journalistic}. On the \emph{micro-level}, we study interactions between hate and counter speakers, including how they support and reply to each other and how each particular utterance changes the overall subsequent discourse~\cite{mathew2019thou}.

\subsection{Macro-Level Effectiveness Measures}

\begin{figure}[t]%
  \includegraphics[width=0.95\textwidth]{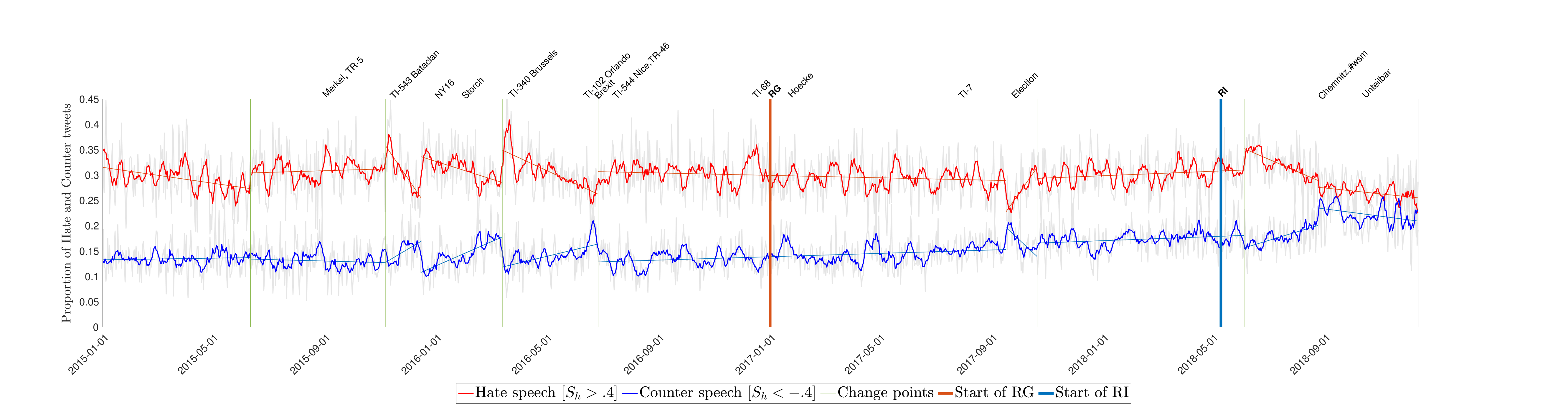}
  \caption{\textbf{Proportion of hate and counter speech over time.} Organized counter speech (RI, blue vertical line) is followed by changes in proportions of hate and counter speech tweeted within 181,370 reply trees to 23 prominent twitter accounts, from January 2015 to December 2018. Each data point is a daily average and trends are smoothed over a one-week window. Exploratory change point analysis \cite{killick2012optimal} identifies the changes in linear trends shown by green vertical lines (see text for further confirmatory statistical tests). Event labels: Merkel = prime minister pronounces `Wir schaffen das' in support of immigrants, TR(TI)$-x$ = right-wing (Islamic) terrorist attacks where $x$ is the number of dead or injured people (location included when attack was outside Germany), NY16 = New Year's Eve assaults on women in Cologne's main train station, Storch = AfD politician calls for use of arms against immigrants at the boarder, Brexit = referendum date, RG = Reconquista Germanica and associated groups start organizing, Hoecke = AfD politician calls for a change in how Germany remembers the Holocaust, Election = German Federal Election, RI = Reconquista Internet becomes functional, Chemnitz = large anti-immigrant protests, Unteilbar - large counter rally in Berlin.}
   \label{fig:macro_measures}%
\end{figure}

Figure~\ref{fig:macro_measures} shows a time series of the proportions of counter and hate speech that occurred in our corpus of reply trees over time. Prior to May of 2018, when RI became functional, the relative proportions of both hate and counter speech are largely in a steady state. Roughly 30\% of the discourse in our sample of political conversations during this time is hateful and only around 13\% was counter speech. However, both of these time series are quite noisy, and there are several clear deviations reflecting the ebbs and flows of political discourse. Many of these deviations coincide with major events such as large-scale terrorist attacks, political rallies or speeches, and elections. After each deviation, however, the proportion of hate and counter speech reverts to the earlier equilibrium, suggesting that these were so-called shock events in an otherwise steady-state system. 

After the formation of RI in early spring of 2018, there was a notable change: a decreasing trend in the proportion of hate speech and an increasing trend in the proportion of counter speech. These trends continued until approximately September 2018, at which point they more or less stabilized. That point coincided with large alt-right rallies in Chemnitz, Germany, followed by a large counter rally in Berlin. Around that time the proportion of hate and counter speech stabilized at similar levels, with counter speech rising to approximately 21-22\% and hate falling to around 25\%. Both proportions continued to decline (although very slightly) through the remainder of the time period studied.

To complement these qualitative observations, we conducted three additional analyses  of the interactions of hate and counter speech: an  exploratory change point detection to detect changes in trends over time, a confirmatory analysis of trends before and after the occurrence of RG and RI, and a vector autoregressive analysis to study the relationship between present and future proportions of hate and counter speech.

First, the exploratory change point detection helped to identify regions of the time series with similar overall trends in proportion of hate and counter speech (see Methods and Refs.~\cite{killick2012optimal,aminikhanghahi2017survey} for details). These regions are generally separated by ``change points'' or important events (e.g., terrorists attacks or political events), which shocked the system and impacted the discourse dynamics (sometimes just temporarily). The automatically identified trends are shown in Figure~\ref{fig:macro_measures} as straight, thin red and blue lines, accompanied by green vertical lines which signify the change points selected by the algorithm. This analysis largely confirmed the descriptive analysis above.  For large portions of time, e.g., from the summer of 2016 to a month before the German federal election in September of 2017, the trends in hate speech and counter speech were roughly constant. Around the time of the German federal election there is a lot of volatility, with major influxes of counter and hate speech. This reflects the political debate between mainstream and far right political views, with the sharp rise in hate speech leading up to the elections being in line with RG's stated goal of swaying the 2017 election toward the AfD, a so called ``radical right” (``rechtsradikal") party. After the election and until the formation of RI there is again a constant trend in the proportion of hate and counter speech. Following the formation of RI, there is a noticeable increase in the proportion of counter speech and decrease in the proportion of hate speech, with these trends stabilizing around the time of Chemnitz and Berlin rallies. The analysis also identifies several other periods with decreasing hate and increasing counter speech. However, these periods were effectively relaxation periods following large shock events such as the Islamic terrorist attacks in Paris and Brussels, which both caused an increase in hate speech and then a relaxation as these events faded from the public focus. 

Second, confirmatory statistical analysis helped us to estimate the impact of RG and RI on hate and counter speech trends. We fit linear trends with slope $m$ to  time series of 1, 3 and 6 months taken before and after the formation of RG and RI, and analyzed the differences in the before and after slopes. If RG had an effect, we would expect a positive change in slope of hate speech trends, i.e., a positive $\Delta_m = m_{\mathrm{after}}-m_{\mathrm{before}}$ around the formation of RG. We would 
expect similar results for counter speech trends around the formation of RI. We would also expect negative differences in slopes for hate speech around the formation of RI, but not necessarily vice versa because organized counter speech was not yet developed around the time when RG formed. The difference in slopes for the proportion of hate was indeed positive 1 month after formation of RG ($\Delta_m = .003$, $CI = (.002, .004)$), and it remained so when comparing slopes 3 and 6 months before and after RG. However, there was a strong increase and decrease in hate speech in the months before the formation of RG, 
linked to a large Islamic terrorist attack that may be biasing this analysis. The change in slope $\Delta_m$ for counter speech before and after RG was less consistent: $\Delta_m$ was slightly positive 1 month after the occurrence of RG ($\Delta_m=.0005$, $CI=(-.0003, .001)$) but decreased 3 and 6 months later ($-.00008$, $CI=(-.0002 -.00002)$). During this time there was no organized counter speech, so this could suggest an initial spontaneous backlash to the emergence of RG that faded after 3 to 6 months. 

After the formation of RI, $\Delta_m$ for counter speech was indeed positive both when comparing slopes 1 month before and after ($\Delta_m=.001$, $CI=(.0005 .002)$) as well as 6 months before and after ($\Delta_m=0.0003$, $CI=(.0003 .0004)$), although not for 3 months before and after ($\Delta_m=-.00002$, $CI=(-.0001 .0002)$). When looking at hate proportions around the time of RI, we see a consistently negative $\Delta_m$ suggesting that hate declined in proportion around the time RI became organized on 1- ($\Delta_m=-.0002$), 3- ($\Delta_m=-.0002$), and 6-month ($\Delta_m=-.0005$) scales.

Third, we used Vector Autoregression (VAR) models to study the relationship between the current proportion of hate and counter speech and their future values over time during three different periods: i) before the establishment of RG (January 2015 - end of 2016), ii) after RG but before RI was established (January 2017 - April 2018), and iii) after the establishment of RI (May - December 2018). 

\begin{figure}[t]%
   \centering
   \includegraphics[width=1\textwidth]{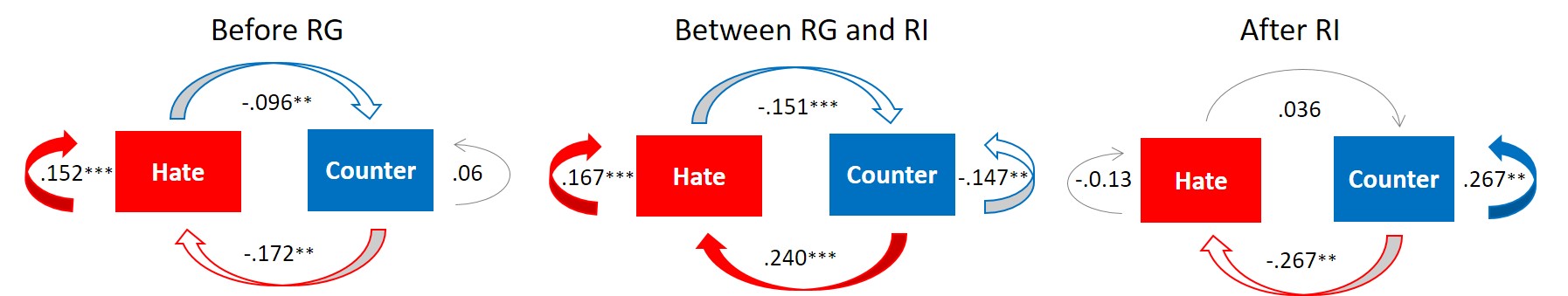}
   \caption{\textbf{Predictive analysis of hate and counter speech.}The relationships between current and future proportions of hate and counter speech for three periods of study. Shown are the results of VAR models for the periods before RG, between RG and RI, and after RI. Red (blue) arrows show statistically reliable effects (** and *** represent $p<0.01$, and $p<0.001$ respectively) on hate (counter) speech, and gray show nonreliable effects ($p>0.05$). Full arrows show positive relationships, and empty arrows negative relationships. Before RG was established, the effects of hate and counter speech were relatively balanced. After the occurrence of RG and before RI, past hate and counter speech became related to more hate speech going forward. After RI was established, past counter speech now negatively predicted future hate and positively predicted future counter speech. }
   \label{fig:vars}%
\end{figure}

Figure~\ref{fig:vars} shows the relationships between current and future proportions of hate and counter speech for the three time periods of study based on the VAR model coefficients. This analysis suggests statistically significant changes in the day to day relationships between the proportion of hate and counter speech after the establishment of RG and RI. When discussing these results we will refer to a ``positive''/``negative'' relationship when the discussed quantity is expected to increase/decrease in the future, respectively. It is important to note that this analysis cannot point to causality of these relationships.

Before RG, the VAR model suggests that there was a positive relationship between the proportion of hate speech and itself, i.e., the presence of hate speech today would suggest a higher proportion of hate the next day. However, during this time there was a negative relationship between counter and hate speech, i.e., more counter speech today would mean less hate speech tomorrow---acting as a dampening factor on hate speech. Also during this time, hate had a negative relationship with counter---more hate speech today implied less counter speech the next day. As discussed in the Introduction, both of these trends might have been a consequence of the fact that both hate and counter speech were conducted by isolated individuals, who might have been intimidated by the opposition they received. Finally, during this period, counter speech did not have a significant relationship with itself, i.e., more or less counter speech today did not imply anything about the future values of counter speech.
After RG was established but before RI, counter speech no longer seemed to dampen hate speech and instead dampened counter speech. In particular, the presence of counter speech on one day would result in \emph{more} hate speech and less counter speech the subsequent days---perhaps suggesting that hate speakers were organizing to target counter speakers, while counter speakers, who were still acting as individuals, were discouraged by the hate they received. During this time, hate speech also had a positive relationship with itself. All of these relationships changed after RI was established, again as expected theoretically given that counter speakers now had a much stronger peer support. In particular, counter speech led to more counter speech while also dampening hate speech and hate speech no longer had a significant association with either future hate or counter speech.

Taken together, the macro level analysis suggests that the emergence of the organized counter speech group RI might have pushed the German political discourse on Twitter into a new state, with a more balanced presence of both hate and counter speech.

\subsection{Micro-Level Effectiveness measures}

However, the effectiveness of this counter speech group cannot be studied in a societal vacuum. The time following the formation of RI was characterized by large political rallies and extensive discussions on both sides of the political spectrum, and  it is not possible to make causal inferences about the effects of either organized hate or counter groups on Twitter. Therefore, we continue with micro level measures of effectiveness to understand the associated changes on a more nuanced descriptive level. 

We begin by analyzing the support that hate and counter speech receives through likes, subsequent replies, and retweets. We then investigate how each hate and counter tweet steers the subsequent discussion in reply trees towards more of the same or towards opposing speech. 

\noindent\textbf{Showing support}

\begin{figure}[ht]%
   \centering
   \includegraphics[width=1\textwidth]{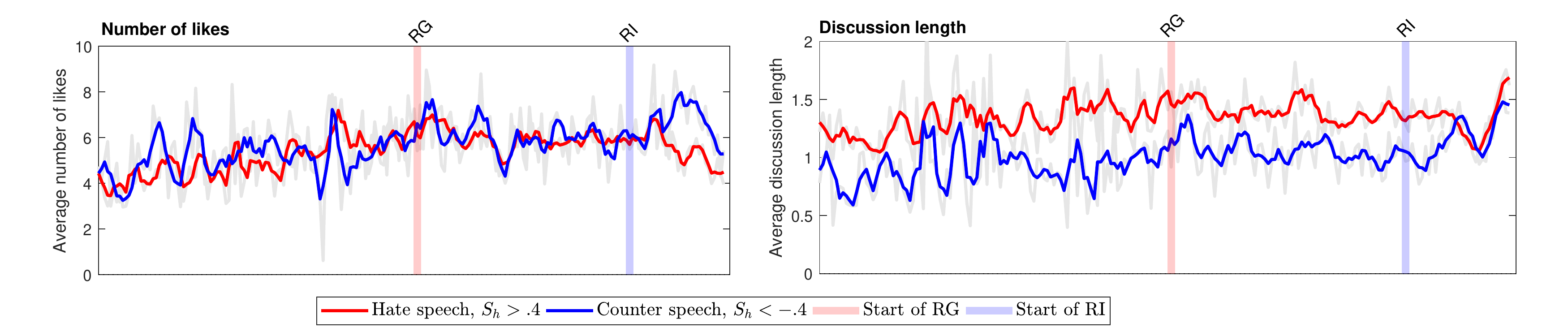}
   \caption{\textbf{Impressions of hate and counter speech.}Impact of hate and counter speech messages over time as quantified by the average number of likes and length of conversation they initiate. The emergence of organized counter speech (RI, blue vertical line). Results are for 181,370 reply trees from January 2015 to December 2018. Each data point is a week average and trends are smoothed over a month-long window. The timeline on the $x$-axis is the same as in other figures but was omitted for space, except for markers of the emergence of RG and RI. 
   } %
   \label{fig:impressions}%
\end{figure}

At the most basic level, participants in Twitter conversations can show their support (or lack thereof) for tweets by affording them their likes and engaging in discussions with them. Engaging in discussion however is not necessarily a sign of support and could be voicing dissent. Figure~\ref{fig:impressions} shows that throughout most of the studied period hate and counter speech received a similar number of likes, but that hate tweets tended to attract longer discussions (measured as the total number of messages in the subtree that a tweet initiated). However, after the emergence of organized counter speech, counter tweets started receiving more likes and attracting longer discussions. The emergence of organized hate speech was not associated with similar changes.

\noindent\textbf{Steering a conversation}

A more fine-grained measure of effectiveness is whether individual tweets can steer the subsequent conversation at the level of a single discussion item.  We calculate the change in average discourse after a tweet is posted in a reply tree, and measure the impact $I$ of a tweet as the difference between the average hate score $S_h$ (defined in Eq.~\ref{eq:hate_score}) of all tweets following and preceding it. More formally, let $t_i$ be the $i^{th}$ tweet that occurred in a tree with $N$ total tweets. Then the impact, or the ability for a tweet $t_i$  to steer a conversations direction, is defined as:
\begin{equation}
\label{eq:impact}
I(t_i) = \frac{1}{i-2}\sum_{j=1}^{i-1}S_h(t_j) - \frac{1}{N-(i+1)}\sum_{j=i+1}^{N}S_h(t_j),
\end{equation}
where $S_h(t_j)$ is the hate score associated with the $j^{th}$ tweet, $t_j$. We may refer to $t_i$ as a ``focal'' tweet when convenient as it is the focus of the computation.  We correct the impacts for factors that could have affected it, e.g., discussion length or average hate-score, using a longitudinal mixed linear model (see Methods and Table \ref{tab:model_diffs} for details). 

\begin{figure}[t]%
   \centering
      \includegraphics[width=0.95\textwidth]{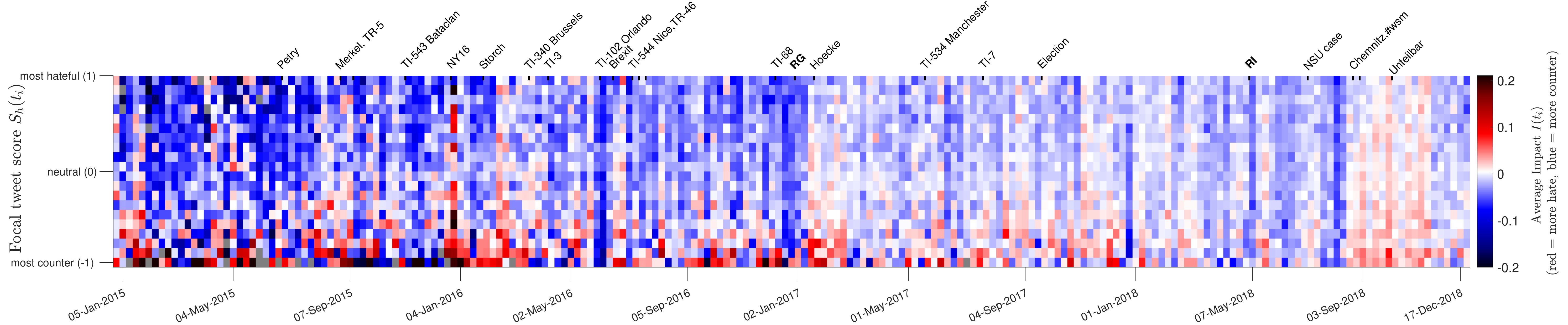} 
   \caption{\textbf{Impact of hate and counter focal tweets on subsequent discourse over time.}
   Each colored square represents the average difference between tweets following and  preceding a focal tweet posted in a reply tree in a particular week (Eq.~\ref{eq:impact}).
   Results are computed for 82,132 reply trees with at least 3 tweets, over time (horizontal axis), and are binned by hate-score of the focal tweet (vertical axis). Red (blue) squares signify that the focal tweet was followed by a change in the subsequent conversation towards more hateful (counter) speech, with color saturation corresponding to the size of the difference. Event labels are the same as in Figure~\ref{fig:macro_measures}.
   }%
   \label{fig:heatmap}%
\end{figure}

Figure~\ref{fig:heatmap} shows the results of this analysis. Time is binned into one week segments and shown on the horizontal axis, while the impact of focal tweets are shown on the vertical axis and binned based on their score, going from  -1 (most counter) to 1 (most hate).
Each square represents the average impact $I(t_i)$ of all tweets occurring in that week with $S_h(t_i)\in[x-0.05,x+0.05]$. This plot allows us to study the average ability for tweets with a given hate score to steer conversations in that week. For this analysis, we needed to ensure that a full-blown conversation occurred, so we restricted the sample of reply trees to 82,132 trees that contained at least three tweets (the initial or root tweet, and at least two replies), for a total of 943,822 tweets.

 The results shown in Figure~\ref{fig:heatmap} suggest that, in general, hateful tweets 
 were followed by counter speech no matter of their hate scores. Counter tweets 
 tended to be followed by hate speech if they had very low hate scores, but by more counter speech when they had moderately low scores. Analyses using the longitudinal mixed linear model mentioned above further suggested that tweets near the end of a tree were especially likely to attract more opposition  (Table \ref{tab:model_diffs}). Tweets in larger trees, and in trees that were already biased towards a particular kind of speech, tended to receive more support. 
 
Throughout most of the studied period, political conversations tended to include somewhat more hate than counter speech (Figure~\ref{fig:macro_measures}), but overall  contained a large amount of neutral speech (average hate score across all trees was $S_h=0.14\pm0.048$). However, the large blue areas in Figure~\ref{fig:heatmap} indicate that as conversations progressed, the average scores of the conversations shifted towards more neutral or even the counter side of the speech spectrum. There were only a few exceptional weeks in which discussions tended to be steered toward hate no matter the score of the focal tweet. An example is the week of 2016's New Year's Eve, when numerous sexual assaults on women in Cologne's main train station were blamed on Syrian and North African refugees and asylum seekers. 
 
There was a notable difference in discourse dynamics from January 2017 onward, as seen by the transition visible around that time in Figure~\ref{fig:heatmap}. This transition corresponds to the time when RG began organizing. Unlike before, hateful conversations in this period did not drift towards counter speech, but stayed neutral or even drifted towards more hateful discourse. Of note, this transition cannot be explained by the classifier picking up on speech not present before 2017 because, as can be seen in Figure~\ref{fig:macro_measures}, the overall proportion of hate speech, as identified by the classifier, is stable around that time. Training data selection bias is also unlikely to have played a role, as shown by Figure~\ref{fig:tweets_year}. The shift in dynamics is also not explained by any known change in the Twitter interface or algorithms\footnote{\url{https://github.com/igorbrigadir/twitter-history/blob/master/changes.csv}}. 
 
 Another transition occurred around the time when RI started to organize in the Spring of 2018. Counter speech started to again dampen hate and pull conversations away from hateful rhetoric. This period was followed by several months of backlash during which almost all tweets were associated with more subsequent hate. The backlash likely reflected the broader societal situation in the Summer of 2018 when large alt-right rallies took place throughout Germany, e.g., in Chemnitz. 
Finally, the Fall of 2018 was characterized by a return of a more effective counter speech, likely bolstered by large counter rallies occurring in October of 2018, such as ``Unteilbar" in Berlin.

\noindent\textbf{Unpacking the pivots}

To understand the dynamics unfolding in Figure~\ref{fig:heatmap} in more detail, we unpack how focal tweets pivot the conversations towards either type of speech. We study four possible reactions to each focal tweet: the subsequent discourse can appear to be \emph{supporting} the focal tweet (i.e., counter focal tweet is followed by more counter speech, and vice versa for hate), \emph{opposing} the focal tweet (counter focal tweet is followed by more hate speech), \emph{polarizing} whereby previously neutral discourse becomes more similar to hate or counter speech after the focal tweet, or \emph{ignoring} the focal tweet (discourse remains unchanged before and after the focal tweet). This analysis also reveals the reasons for changes in macro level indicators of hate and counter speech shown in  Figure~\ref{fig:macro_measures}. 

\begin{figure}%
   \centering
   \includegraphics[width=1\textwidth]{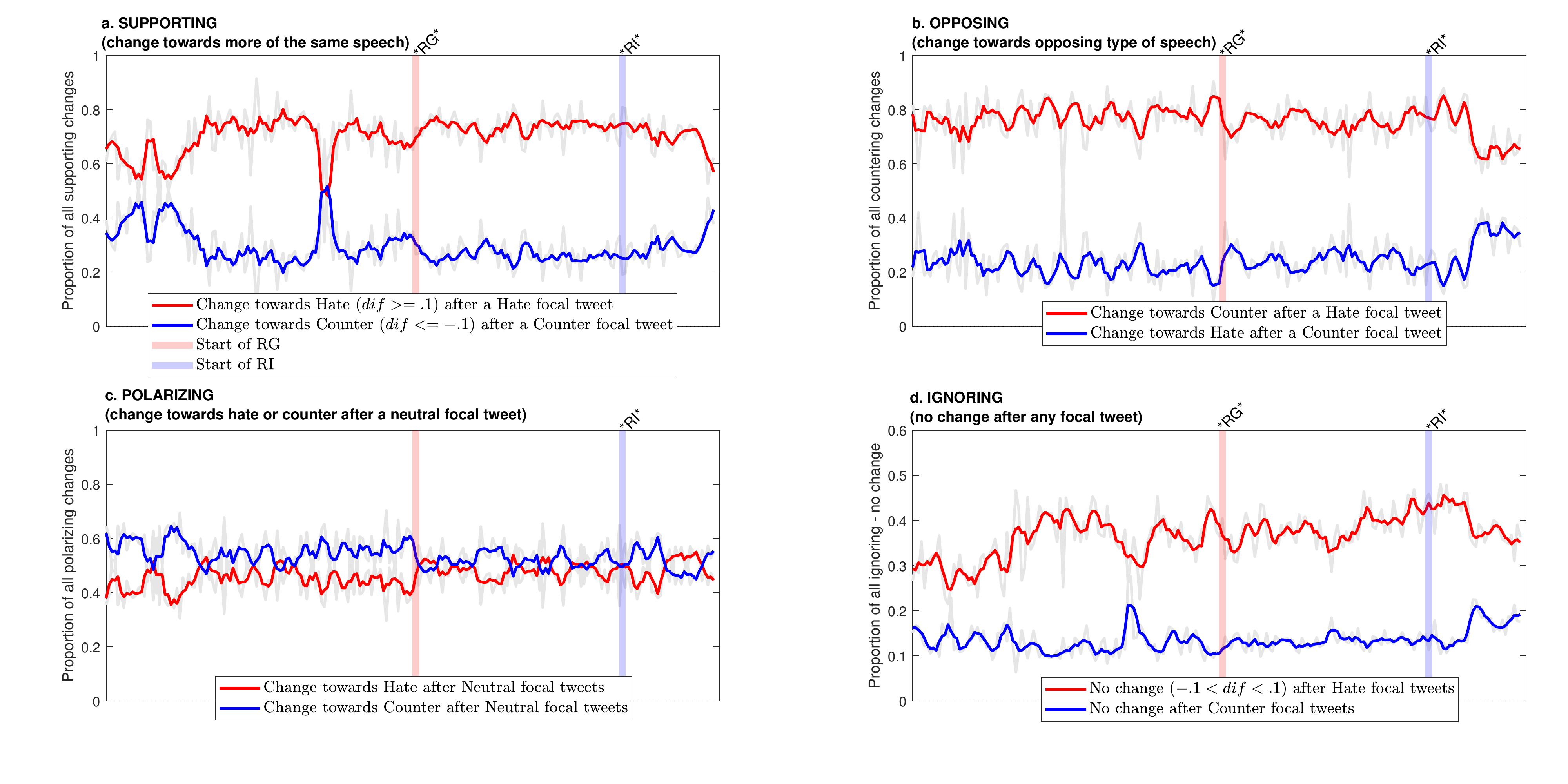}
   \caption{\textbf{Reactions to tweets of different types.}.Proportion of focal tweets followed by different reactions, for 82,132 reply trees with at least 3 tweets. Shown are: numbers of focal tweets each week that are followed by a) supporting, or changes towards the same type of speech, b) countering, or changes towards the opposing type of speech, c) polarizing, or changes towards hate or counter after a neutral focal tweet, and d) ignoring the focal tweet. Each data point is a week average and trends are smoothed over a month-long window. 
   }%
   \label{fig:heatmap_unpack_prop}%
\end{figure}

Figure~\ref{fig:heatmap_unpack_prop} shows the relative proportion of different types of reactions that occurred after hate, counter, or neutral focal tweets. In general, hate focal tweets were more often followed by supporting, opposing, or ignoring reactions than counter speech. These trends were relatively stable but could be temporarily altered by notable events, such as the attack on an Orlando night club in June 2016, when counter speech received a burst of support. These results suggest that the notable change in discourse dynamics from January 2017 onward, visible in Figure ~\ref{fig:heatmap}, occurred primarily because neutral speech became more polarized in the direction of hate speech around that time (panel c.). In other words, neutral reply trees started to be steered towards more hateful speech. In addition, hate speech became a bit more supported around that time (panel a.).

This analysis also sheds light on the reasons for relative decrease in frequency of hate speech after the emergence of RI (Fig.~\ref{fig:macro_measures}), as well as for the fact that conversations  were more often steered towards counter speech around that time  (Fig.~\ref{fig:heatmap}). Figure~\ref{fig:heatmap_unpack_prop} suggests that this occurred because of an increase in support for counter tweets (panel a.) and, for a short while, an increased polarization of neutral speech towards counter speech (panel c.). However, as suggested in (Fig.~\ref{fig:heatmap}), a period of hate speech backlash followed the early successes of RI. According to Figure~\ref{fig:heatmap_unpack_prop}, this occurred mostly because opposing discourse became more frequent after counter speech than after hate speech (panel b.), and neutral speech became more polarized towards hate again (panel c.).  

\section{Discussion}
\label{sec:discussion_nhb}

We present a large-scale longitudinal study of the dynamics of hate and counter speech in political discussions online. By analyzing hundreds of thousands of Twitter conversations through different lenses at both macro and micro levels, we contribute a nuanced picture of these dynamics. Across a number of different indicators, we find that organized counter speech appears to contribute to a more balanced public discourse. After the emergence of the organized counter group Reconquista Internet (RI) in the late Spring of 2018, the relative frequency of counter speech increased while that of hate speech decreased (Fig.~\ref{fig:macro_measures}). Counter speech became more related to less hate and more counter speech in the future (Fig.~\ref{fig:vars}). The number of likes and the length of discussions after counter tweets increased after RI was founded (Fig.~\ref{fig:impressions}). Counter speech became more effective in steering conversations when it organized through RI, primarily by providing more support to counter tweets and by steering neutral discourse towards more counter speech (Figs.~\ref{fig:heatmap} and~\ref{fig:heatmap_unpack_prop}). While this was met with a backlash initially, the relative frequency of hate speech stabilized into a new, lower proportion of overall speech (Fig.~\ref{fig:macro_measures}). These findings suggest that, like in `traditional' bullying settings, the presence of supporting peers (in this case, other individuals willing to engage in counter speech) motivates people to themselves oppose hate speech and defend its targets.

According to our results, citizens wishing to engage in counter speech would likely increase the effectiveness of their efforts if they organized and participated in discussions in a coordinated way. 
 For example, they could organize around a central platform where members can communicate and strategize  (e.g., RI members communicated via Discord server).  
As an organization they could then steer hateful conversations to a more neutral or even positive ground, by supporting the victims, voicing dissent to hateful positions, or even simply by ``liking'' counter messages so that they are more broadly visible. This is in line with similar results that were found qualitatively in \cite{cathyCounter2020}. 

Although we see an association of hate and counter speech and subsequent discourse according to a number of indicators, we cannot draw causal conclusions about the effectiveness of organized hate or counter speech alone. German society has been undergoing significant self-examination about its values and political directions throughout the studied period and beyond. This has likely influenced the ongoing discourse beyond organized counter speech. Regardless, the multifaceted approach we took here suggests that organized counter speech may be a promising strategy in combating the spread of hate online.

There are several critical areas of future research on counter speech\footnote{For a comprehensive review of past research on counter speech see \cite{cathyCounterReview2019}.}. One is the analysis of which counter speech strategies are particularly effective in curbing hate. A few recent studies have begun to look at this problem qualitatively. For example, \cite{cathyCounter2020} interviewed more than 25 active counter speakers who were members of the so-called ``I am here" (\#jagärhär) counter speech movement. Asking important questions like, why they took part, what the counter speaker learned or got out of the experience, and perhaps most importantly what strategies did they find effective or detrimental in combating online hate. This kind of work is crucial because it allows for a first-hand account of the strategies being used but unfortunately does not lend itself to any quantitative analysis on these speakers strategies. Additionally, Keller \& Askanius~\cite{keller2020combatting} used the same situation between RG and RI to perform a qualitative content analysis to understand the strategies being used by both groups. To this end, they examined the internal documents of both groups as well as 709 comments generated in response to 3 articles posted on Facebook. They concluded that ``organized, communication- and exchange-oriented, rule-guided efforts can be effective to combat cyberhate,'' but also state that in some circumstances counter speech can cause further polarization of beliefs. Again, this analysis did not allow for any causal analysis of the effects of either group. An exciting avenue for future research would be to combine the tools and lessons of these qualitative studies with the large corpus we have used here to drill into and identify which of the many strategies being used by these groups was effective.  

Another important open question is how events such as terrorist attacks and political rallies shape hate and counter speech trends. In our results, we observed some patterns echoing previous results on the relationship between group threat and extreme attitudes and behaviors \cite{jahnke2021violence,jost2003political,turner1992threat}, but without a clear baseline we cannot derive any strong causal conclusions. Future research could try to find possibilities for including exogenous events that could enable such analyses \cite{muller2019fanning}. 

\new{As it is common for users to report accounts of the opposing faction for policy violations, yet another open question is how hate and counter speech relate to longevity of user accounts, in particular what percent of different accounts survive over time. Unfortunately, out data does not allow for these analyses \emph{post hoc} as it is impossible to say whether an account was deleted by the user, suspended temporarily or permanently for policy violations. Our data also does not allow us to detect ``shadow banning", periods when an account appeared to be active but its posts were not visible to others. Future research could try to record these indicators in real time or with the help of the platform provider. Similarly, it would be interesting to study the percentage of removed tweets that were flagged as either hate or counter by our classification algorithm. But, again drawing any conclusions about \emph{why} (e.g., removed for policy violations, deleted by user, originated from suspended account, etc.) a tweet was removed would be challenging but could provide insight into the longevity of hate and counter speech on online platforms.  
}  

\new{Recently, there has been a lot of interesting work (e.g., \cite{alizadeh2020content,pacheco2021Uncovering,pacheco2020unveiling}) on detecting coordinated activity on social media, especially around propagation of misinformation. While this work is adjacent to our work it actually solves the converse problem. These studies focus on detecting suspected coordinated activity, while we already knew there was coordinated activities and were interested in studying the dynamics of these activities in our dataset. However, given the organized nature of these groups, it would be interesting to study whether automation such as bots or manipulation of trending topics were used by these groups. In the present work, we were interested in the nature and dynamics of this discourse no matter how they were generated i.e., via bot or human. However, future research should certainly adapt tools such as Botometer~\cite{davis2016botornot,varol2017online}, Botslayer~\cite{hui2019botslayer} 
, and similar tools to be applicable to Twitter reply trees. These analyses would provide valuable information on the mechanics of coordinated activities online.}

\new{In August of 2020 Twitter announced that in the second iteration of the Twitter API, conversations which occurred in the previous week are now obtainable directly from the API using the new ``conversation\_id" query feature. In January of 2021, the one-week restriction on conversation queries was relaxed when Twitter announced the ``Academic Research product track" (or Academic API). With it, academics are now granted the ability to perform full-archival search of Twitter's history using any of the API's many endpoints and search queries. Hence, going forward academic researchers will be able to do longitudinal analysis similar to that which we are reporting here. However, there are a few caveats which we feel are worth briefly discussing. First, at the time of writing this, the full-archival search functionality was still in early access and many older reply trees were still unsearchable. During revisions of this paper we attempted to reconstruct our conversation dataset using this new API but were unsuccessful. In our experience, as of August 31, 2021, only 0.02\% of the requested trees were partially returned and the rest were not found (see Section~\ref{sup:academicAPI} for a full discussion of this analysis).  Second, in recent years Twitter has worked to remove content from their platform which violated its Terms of Service. While this is beneficial to the Twitter community at large, this means that when researchers query the new API for older conversations, the API's response may have a significant portion of the more extreme tweets removed. This effectively washes out the signal that was of most interest in the present research. See Figure~\ref{fig:susp} for an example. In this figure, half of the conversation has been removed because the person participating in the conversation has been suspended. In other conversations we explored (not shown), we see similar messages involving tweets removed by users, removed for violation of Terms of Service etc.  As such, while full-archival search of conversations will be possible going forward, compiling Twitter conversations in real time (as much as possible) will continue to be necessary when studying behavior that violates the platform's Terms of Service such as hate speech.} We hope this study provides a template for future longitudinal studies of conversation dynamics.

\section*{Competing interests}
  The authors declare that they have no competing interests.

\section*{Author's contributions}
   JG and KG-Z designed the study, collected/scraped the Twitter data used in this study, as well as developed the classification pipeline. All authors contributed to analysis and interpretation of the data and to the writing of the manuscript. All authors read and approved the final manuscript.

\section*{Acknowledgements}
     J.G. was partially supported by an Omidyar and an Applied Complexity Fellowship at the Santa Fe Institute as well as NSF EAGER 1807478.
   J.-G.Y. was supported by a James S. McDonnell Foundation Postdoctoral Fellowship Award.
   L.H.-D. and J-G.Y. were supported by Google Open Source under the Open-Source Complex Ecosystems And Networks (OCEAN) project. M.G. was partially supported by NSF-DRMS 1757211. 
   Any opinions, findings, and conclusions or recommendations expressed in this material are those of the authors and do not necessarily reflect the views of Google Open Source or the NSF. \new{Finally, the authors would like to thank the anonymous reviewers for their valuable feedback.}

\bibliographystyle{unsrt} 
\bibliography{references}      
\newpage

\beginsupplement
\renewcommand{\thesection}{S\arabic{section}}  
\renewcommand{\thetable}{S\arabic{table}}  
\renewcommand{\thefigure}{S\arabic{figure}}
\setcounter{section}{0}
\section{Supplementary Information}
\label{sup:methods}
\subsection{Twitter's Academic Track API Experiment}
\label{sup:academicAPI}

\new{During revisions to this manuscript Twitter announced the academic API which allows for ``complete archival search" of Twitter conversations using the conversation\_id query. During August of 2021, in an attempt to verify our reply tree dataset, we tried to recollect a sample of all of our trees using the full-archive search v2 endpoint of the academic API. To this end, we requested 47,297 of our conversations (approximately 25\% of our dataset) from the API using the conversation\_id query functionality. Of these requests only 9 trees were actually found by the API, the other 47,282 requests returned an empty json response from the API, indicating the query had no results. Of the 9 trees that were found and returned by the API, all of them were only partial returns, i.e., when we compared the API's full json response (even after carefully taking into account pagination and similar technical issues) to the conversation that is available online the majority of replies were missing (e.g., in one conversation we only received back 15 of the 2,464 possible replies).}

\new{In summary, while some older reply trees/conversations were searchable using the conversation\_id search features of the API, the majority of older conversations were not found at all or only returned a small subset of the conversation. In our experience as of August of 2021 only 0.02\% of our requested trees were partially returned and the rest were not found. Hence, it seems that while Twitter is working towards allowing full-archival search with the academic API, the Twitter database has not been completely updated to allow for full functionality.
}

\begin{figure}
   \centering
   \includegraphics[width=.5\textwidth]{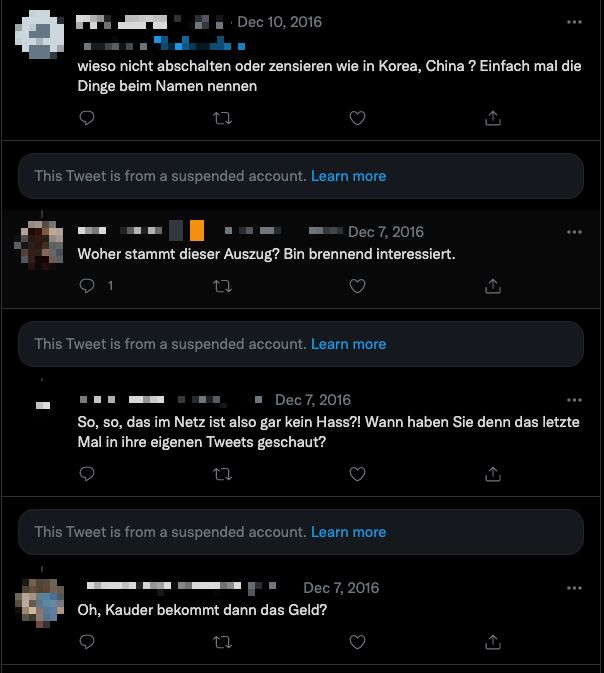}
   \caption{An example where tweets available from the API are missing because the tweet author was suspended, effectively washing out the signal of interest.}
   \label{fig:susp}
\end{figure}

\new{An additional concern related to doing this kind of historical collection of conversations when studying hate speech is that the API does not return tweets from suspended accounts, tweets removed by users, or tweets that have been removed for violating terms of service. This, of course, occurs a lot around hate speech and means that a significant portion of the signal we are interested in has been washed out of the older conversations we can query. For example, Figure~\ref{fig:susp} shows a segment of a conversation of which half the replies have been removed from Twitter, and thus the API response, because the person participating in the conversation has been suspended.}

\subsection{Supplementary Figures}

\begin{figure}[ht]%
   \centering
   \includegraphics[width=0.9\textwidth]{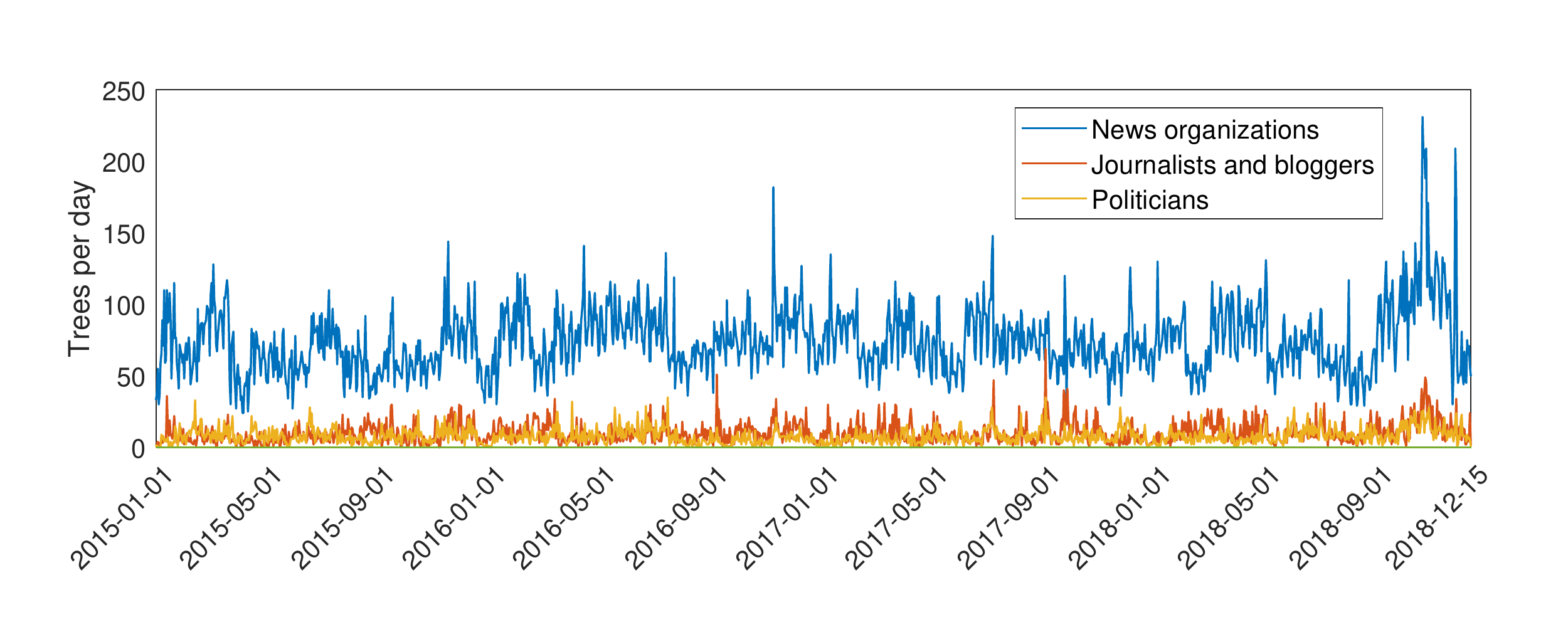}%
   \caption{Number of reply trees per day, by type of account, for a total of 181,370 trees used in analyses.}. 
   \label{fig:trees_day}%
\end{figure}

\begin{figure}
   \includegraphics[width=0.9\textwidth]{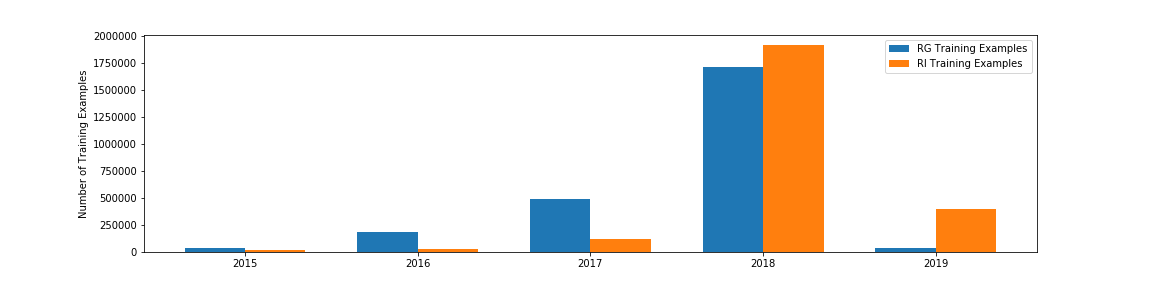}
   \caption{Number of tweets used as training examples for the classification pipeline, for each class per year.}.
   \label{fig:tweets_year}
\end{figure}

\begin{figure}%
   \centering
\includegraphics[width=1\textwidth]{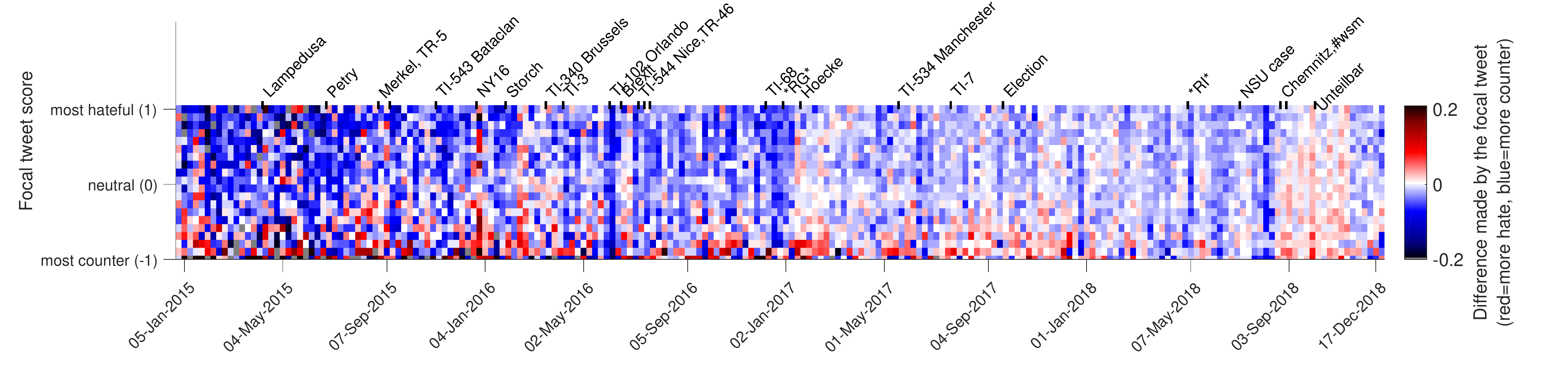}%
\\
\includegraphics[width=1\textwidth]{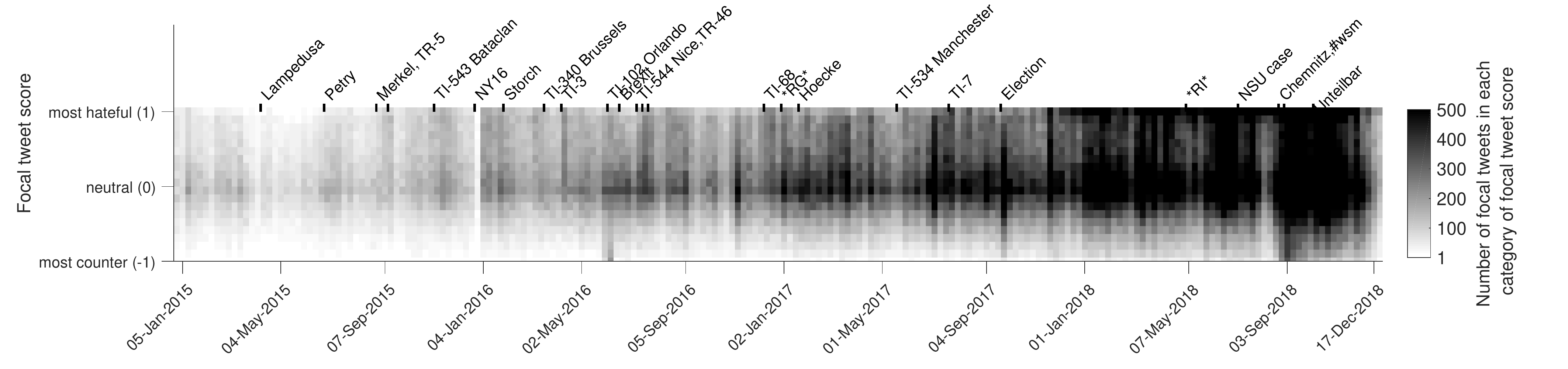} %
   \caption{Top panel: Uncorrected results for Fig.~\ref{fig:heatmap}. Bottom panel: Number of focal tweets in each category of focal tweet score.}%
   \label{fig:heatmap_raw}%
\end{figure}

\newpage

\subsection{Supplementary Tables}

\begin{table}[]
\centering
   \begin{tabular}{l|ccc}
   \hline
    & \textbf{Estimate} & \textbf{Lower Bound} & \textbf{Upper Bound} \\
    \hline
   \textbf{Fixed effects}                         &         &         &         \\
   Intercept                             & -0.0994 & -0.1179 & -0.0810 \\
   $S_h(t_i)$                            & -0.0182 & -0.0265 & -0.0099 \\
   Focal tweet  position $\mathrm{pos}(t_i)$      & 0.0001  & 0.0001  & 0.0002  \\
   Tree size                             & 0.0002  & 0.0001  & 0.0003  \\
   Average tree hate score               & -0.0097 & -0.0311 & 0.0116  \\
   Week $w$                              & 0.0002  & 0.0002  & 0.0003  \\
   Politicians vs. news outlets          & 0.0690  & 0.0597  & 0.0784  \\
   Journalists/bloggers vs. news outlets & 0.0846  & 0.0758  & 0.0934  \\
   $S_h(t_i)$ x  $\mathrm{pos}(t_i)$              & -0.0005 & -0.0005 & -0.0005 \\
   $S_h(t_i)$ x Tree Size                & 0.0002  & 0.0002  & 0.0003  \\
   $S_h(t_i)$ x Average tree hate score  & 0.0165  & 0.0058  & 0.0273  \\
   $S_h(t_i)$ x $w$                      & 0.0000  & 0.0000  & 0.0001  \\
   $S_h(t_i)$ x Politicians              & 0.0010  & -0.0021 & 0.0041  \\
   $S_h(t_i)$ x Journalists              & 0.0009  & -0.0023 & 0.0041  \\
   \hline
   \textbf{Random effects}                        &         &         &         \\
   Tree                                  & 0.3515  & 0.3495  & 0.3536  \\
   Week                                  & 0.0152  & 0.0117  & 0.0196 \\
   \hline
   \end{tabular}
    \vspace{10pt}
   \caption{ Longitudinal mixed linear model of differences due to focal tweets, implemented by Matlab fitlme procedure. More complicated models did not produce better fit indices. 
   The results suggest that authors of both hate and counter speech tend to show clustering behavior. Focal tweets in trees that are overall biased towards the same type of speech, larger trees, and trees started by politicians and journalists as opposed to official news outlets, tend to attract more of the same speech. In other words, in such trees hate was somewhat more likely to attract hate, and counter to attract counter speech. Accounts of politicians and journalists were overall more likely to receive hateful responses compared to traditional news outlets. Tweets near the end of the tree tend to be followed by more opposition (which in turn possibly contributes to ending of the discussion). }%
   \label{tab:model_diffs}%
   \end{table}

\end{document}